\documentstyle[preprint,eqsecnum,aps]{revtex}\tighten

\newcommand{\beq}{\begin{equation}}
\newcommand{\beqa}{\begin{eqnarray}}
\newcommand{\eeq}{\end{equation}}
\newcommand{\eeqa}{\end{eqnarray}}
\newcommand{\be}{\begin{equation}}
\newcommand{\ba}{\begin{eqnarray}}
\newcommand{\ee}{\end{equation}}
\newcommand{\ea}{\end{eqnarray}}
\newcommand{\simg}{\gtrsim}
\newcommand{\siml}{\lesssim}

\begin{document}
\draft

\preprint{KUNS-1493, YITP-98-13}

\title{ Black Hole Binary Formation in the Expanding Universe\\
 --- {\it Three Body Problem Approximation} ---
}

\author{
Kunihito Ioka$^{1}$, 
Takeshi Chiba$^{2}$, Takahiro Tanaka$^{3}$,
and Takashi Nakamura$^{4}$
}
\address{$^{1}$ Department of Physics, Kyoto University, Kyoto 606-01,
Japan}
\address{
$^{2}$Department of
Physics, University of Tokyo, Tokyo 113-0033, Japan
}
\address{$^{3}$Department of Earth and Space Science, Osaka 
  University, Toyonaka 560, Japan }
\address{$^{4}$Yukawa Institute for Theoretical Physics, Kyoto University, 
Kyoto 606-01, Japan}

\date{\today}

\maketitle

\begin{abstract}
We study black hole MACHO binary formation through three-body
interactions in the early universe at $t\sim 10^{-5}$s .
The probability distribution functions of the eccentricity and the
semimajor axis of binaries  as well as of the coalescence time are
obtained assuming that the black holes are randomly formed in space.
We confirm that the  previous order-of-magnitude estimate for the binary
parameters is valid within $\sim$ 50\% error. We find  
that the coalescence rate of the black hole MACHO 
binaries is $\sim 5 \times 10^{-2} \times 2^{\pm 1}$ events/year/galaxy 
taking into consideration
several possible factors which may affect this estimate.
This suggests that the event rate of coalescing 
binary black holes will be at least several events per year within $15$ Mpc.
The first LIGO/VIRGO interferometers in 2001 will be able to verify
whether the MACHOs are black holes or not.
\end{abstract}


\section{introduction}
The analysis of the first $2.1$ years of photometry of $8.5$ million
stars in the Large Magellanic Cloud (LMC) by the MACHO collaboration
\cite{alcock} suggests that the fraction $0.62^{+0.3}_{-0.2}$ of the halo consists of 
MACHOs (MAssive Compact Halo Objects) of mass $0.5^{+0.3}_{-0.2}
M_{\odot}$ in the standard spherical flat rotation halo model.
The preliminary analysis of four years of data suggests the existence of 
at least four additional microlensing events with $t_{dur}\sim 90$ days
in the direction of the LMC \cite{pratt}.
At present, we do not know what MACHOs are. This is especially 
because there are strong degeneracies in any microlensing measurements;
the mass, velocity and distance of a lens object.
The inferred mass is just the mass of red dwarfs.
However, a tight constraint is obtained from the observations
\cite{bahcall,flynn,graff}.
The red dwarfs contribute at most a few percent to the mass of the halo.

The brown dwarfs are also restricted by the {\it Hubble\ Space\
  Telescope} search.
Extrapolating the mass function, the contribution of the brown dwarfs to 
the mass of the halo is less than a few percent\cite{graff}.
However, the possibility of brown dwarfs cannot be rejected
if the mass function has a peak at the brown dwarfs
since they are so dim.
The possibility that the MACHOs are neutron stars is ruled out
by the observational constraints on the metal and helium abundance
\cite{ryu}.

As for white dwarfs, assuming the Salpeter initial mass function (IMF)
with an upper and a lower mass cutoff,
the mass fraction of the white dwarfs in the halo
should be less than $10$ \% from  the number count of the high
$z$ galaxies\cite{charlot}.
The observation of the chemical yield does not favor
that the MACHOs are white dwarfs\cite{gibson,canal}.
However if the IMF has a peak around $\sim 2 M_{\odot}$,
MACHOs can be white dwarfs
\cite{chabrier,adams,fields}.
Future observations of high-velocity white dwarfs in our solar
neighborhood will make clear whether white dwarf MACHOs 
 exist or not.
Of course, it is still possible that an overdense clump of MACHOs exist
toward the LMC \cite{kanya}. 

If the number of the high-velocity white dwarfs turns out to be large
enough to explain the MACHOs, then star formation theory should
explain why the IMF has a peak at $\sim 2M_{\odot}$.
If it is not, we must consider other possibilities such that 
MACHOs are primordial black holes or boson stars. 
If  MACHOs are black holes, however, 
it seems difficult to verify observationally 
whether the MACHOs are black holes or not.
In fact, electromagnetic radiation from gas accreting to a black hole MACHO
(BHMACHO) is too dim to be observed unless the velocity of BHMACHO is
exceptionally small \cite{fujita}.

Recently, however, Nakamura {\it et\ al}. \cite{bhmacho} proposed the
detectability of the gravitational wave from coalescence of BHMACHO
binaries which are formed through the three body interaction in 
 the early universe at $t\sim 10^{-5}$s. 
The event rate of coalescing BHMACHO binaries  was estimated as $\sim
5\times 10^{-2}$ events/year/galaxy,
which suggests that we can expect several events per year within 15 Mpc.
If this estimate is true, not only can we confirm whether 
the MACHOs are black holes or not, but also we have plenty of  
 sources of gravitational waves. 
However in Ref.\cite{bhmacho} they  made only order-of-magnitude
arguments and there are uncertainties in the estimate of the event rate. 
Especially, the estimate of the semimajor and the semiminor axis of the
binary formed through the three body interaction is not based on
accurate numerical calculations.
In this paper we investigate up to what extent the order-of-magnitude
arguments in Ref.\cite{bhmacho} are valid by calculating numerically the
three body problem in the expanding universe.

In \S 2 we will review the formation scenario of the BHMACHO binaries
 in Ref.\cite{bhmacho}.
In \S 3 we will derive the basic equations of a multi-particle system in the
expanding universe.
In \S 4 we will show the results of the numerical calculations of the
binary formation through the three body interaction in the expanding
universe.
In \S 5 we will obtain the probability distribution functions of the
eccentricity and the semimajor axis of binaries as well as the
coalescence time assuming that the black holes are randomly formed in
space.
We will estimate the event rate of coalescing binaries using this
probability distribution function. 
In \S 6 we will consider several possible factors which may affect
the estimate of the event rate, and we will conclude that the event 
rate is not reduced considerably.
In \S 7 we will consider how BHMACHO binaries evolve
after the binary formation.
\S 8 will be devoted to summary and discussions.

We use units of $G=c=1$ in this paper.
\section{review of BHMACHO binaries formation  scenario}

We briefly review the BHMACHO binaries formation scenario in
Ref.\cite{bhmacho} to introduce our notations. 

For simplicity, we assume that black holes dominate the dark matter,
i.e. $\Omega=\Omega_{BHM}$. The extension to the case where   
black holes do not dominate the dark matter is not difficult.
Further we assume that all black holes have the same mass and we
describe it as $M_{BH}$.
(The extension to the unequal mass
case is straightforward and is given in Appendix A.) 
Primordial black holes are formed when the horizon
scale is equal to the Schwarzschild radius of a black hole.
There are some theories about the formation mechanism of primordial
black holes \cite{yokoyama,kawasaki,jedamzik}.
At present, however, we can not say definitely
whether the black holes will form or not in the early universe.
Ultimately, only by observational technique the existence of a
population of primordial black holes may be established.
It is therefore important to establish the observational signatures of
primordial black holes.
Our standpoint is that we are quite ignorant of whether 
large numbers of primordial black holes will form or not in the early universe,
and it is very important to confirm the existence
of primordial black holes observationally. Whether the results of the 
observation confirm the extistence of primordial black holes 
or not, it will add very much to our understanding of the universe.

The scale
factor at the time of the formation is given by
\beq
R_f=\sqrt{ M_{BH} /H^{-1}_{eq}}=1.1\times 10^{-8}\left({M_{BH}\over
M_{\odot}}\right)^{1\over 2}
\left(\Omega h^2\right),
\label{rform}
\eeq
where $H_{eq}$  with  $H^{-1}_{eq}=\sqrt{3/8\pi \rho_{eq}}=  1.2
\times 10^{21}\left(\Omega h^2\right)^{-2}
 {\rm cm}$ is the Hubble parameter at the time of matter-radiation 
equality. We normalize the scale factor  such that $R=1$ at the time of matter-radiation 
equality. 

The mean separation of black holes $\bar{x}$ with mass $M_{BH}$ at the time of 
matter-radiation equality is given by
\beqa
\bar{x}&=&(M_{BH}/\rho_{eq})^{1/3}\nonumber\\
&=&1.2\times 10^{16}(M_{BH}/M_{\odot})^{1/3}(\Omega h^2)^{-4/3}{\rm
cm}.\label{mean}
\eeqa
As a foundation for computing the distribution function of BHMACHO
binaries with respect to the binary parameters, in Ref.\cite{bhmacho}
it was assumed that (i) the BHMACHOs are
created with a distribution of comoving separations $x$ that is
{\it uniform} over the range from an initial physical separation equal to
the black hole size to a maximum separation $x=\bar x$ and that (ii) the
BHMACHOs' initial peculiar velocity
is negligible  compared to the Hubble flow. 
Obviously, it is more realistic to assume that the BHMACHOs
are formed {\it randomly} rather than uniformly. We will  consider 
the effect of the initial peculiar velocity in Section VI.D.

Consider a pair of black holes with the same mass $M_{BH}$ and
a comoving separation $x < \bar{x}$. 
These holes' masses produce a mean energy density over a sphere with
the radius of  the size of their separation as 
$\bar \rho_{BH}\equiv {\rho_{eq}\bar x^3}/({x^3} { R^3})$. 
$\bar \rho_{BH}$ becomes larger than the radiation energy density
$\rho_r ={\rho_{eq}}/{R^4} $ if 
\beq
R>R_m\equiv \left(\frac{x}{\bar x}\right)^3.
\eeq
After $R=R_m$ the binary decouples from the cosmic
expansion and becomes a bound system. 
The tidal force from 
neighboring black holes gives the binary sufficiently large
angular momentum to keep the holes from colliding with each other 
unless $x$ is exceptionally small. 

The semimajor axis $a$ will be proportional to $x R_m$.
Hence, we have
\beq 
a = \alpha x R_m=\alpha{x^4\over \bar x^3}, 
\label{alpha} 
\eeq
where $\alpha$ is a constant of order $O(1)$.
To estimate the tidal torque, we assume that the tidal force is
dominated by the black hole nearest to the binary.
We denote by $y$ the comoving separation of the nearest neighboring 
black hole from the center of mass of the binary. 
Then, from dimensional analysis, the semiminor axis $b$ will be
proportional to (tidal force)$\times$(free fall time)$^2$ and is given
by
\beq
b= {\alpha \beta} \frac{M_{BH}\,xR_m}{(y R_m)^3} 
      \,{(xR_m)^3\over M_{BH}}
    = \beta\left({x\over y}\right)^3 a,
\label{beta}
\eeq
where  $\beta$ is a constant of order $O(1)$.
Hence, the binary's eccentricity $e$  is given by 
\beq
e = \sqrt{1- \beta^2\left({x\over y}\right)^6}.
\label{eccent}
\eeq
In Ref.\cite{bhmacho}, $\alpha=\beta=1$ is assumed. However,
 $\alpha$ and $\beta$ will be different from unity so that
 calculations of the distribution functions based on an accurate
estimate of  $\alpha$ and $\beta$
are necessary . This is the prime subject of the present paper.

\section{multi-particle system in the expanding universe}

Although our main interest is in the three body problem, we formulate
the problem as generally  as possible. 

\subsection{Basic equations}

We treat the motion of a black hole as that of a test particle within
the Newtonian approximation \cite{peebles,futamase,shibata}.
 We first assume that the line element is given by
\beq
ds^2=-(1+2\phi)dt^2+(1-2\phi)R(t)^2d{\bf x}^2, 
\eeq
where $\phi$ is the Newtonian potential determined by
\beq
R^{-2}\Delta \phi =4\pi  \rho_{BH},
\eeq
with $\rho_{BH}$ being the energy density of   black holes.
For the multi-particle system, the potential is readily solved as
$\displaystyle \phi({\bf x}) = -\sum_{\bf j}
{{m_{\bf j}}\over {R |{\bf x}-{\bf x}_{\bf j}|}}$,
where ${\bf x}_{\bf j}$ is the position of the j-th black hole.
The action of the particle is given by 
\beq
\int ds= \int \sqrt{1+2\phi-R^2\dot{{\bf x}}^2}dt \simeq 
\int \left(1-{1\over 2}R^2\dot{{\bf x}}^2 +\phi\right)dt.
\label{lagrangian}
\eeq
Then the equation of motion is derived as
\beq
(R^2 \dot{{\bf x}})^{.}=-\nabla \phi.
\eeq
For the potential, $\displaystyle \phi({\bf x}_{\bf i})
= -\sum_{\bf j \neq \bf i}
{{m_{\bf j}}\over {R |{\bf x}_{\bf i}-{\bf x}_{\bf j}|}}$,
of the multi-particle system, the above equation is
expressed as
\beq
(R^2 \dot{{\bf x}}_{\bf i})^{.}=-{1\over R}\sum_{\bf j \neq \bf i}
{m_{\bf j} 
({\bf x}_{\bf i}-{\bf x}_{\bf j})\over |{\bf x}_{\bf i}-{\bf
x}_{\bf j}|^3}.
\label{eom:multi}
\eeq
We introduce ${\bf z}_{\bf i}\equiv {\bf x}_{\bf i}/\bar{x}$ and 
use the scale factor $R$ as an independent variable. Then 
for the equal-mass black hole case,
Eq.(\ref{eom:multi}) can be written as 
\beqa
{\bf z}_{\bf i}''+{1\over R}{\bf z}_{\bf i}'&=&-{M_{BH}\over \bar{x}^3 
H^2R^5 }\sum_{\bf j \neq \bf i}
{
({\bf z}_{\bf i}-{\bf z}_{\bf j})\over |{\bf z}_{\bf i}-{\bf
z}_{\bf j}|^3}\nonumber\\
&=&-{3\over 8\pi R}\sum_{\bf j \neq \bf i}
{({\bf z}_{\bf i}-{\bf z}_{\bf j})\over |{\bf z}_{\bf i}-{\bf
z}_{\bf j}|^3},\label{eq:z}
\eeqa
where a prime denotes the derivative with respect to $R$ and 
we have used Eq.(\ref{mean}) in the last equality.
Note that Eq.(\ref{eq:z}) does not depend on $M_{BH}$.
Moreover there is  a scaling law, i.e.
Eq.(\ref{eq:z}) is invariant under the transformation defined by
\beq
{\bf z} \to \lambda {\bf z},\quad R \to \lambda^3 R,
\label{scaling}
\eeq
where $\lambda$ is a constant.

\subsection{Validity of Newtonian approximation}

The cosmological Newtonian approximation is valid if (1) $|\phi | \ll 1$ 
and (2) the scale of inhomogeneity $\ell$ satisfies the relation 
 $\ell \ll H^{-1}$
\cite{futamase,shibata}.
Since the minimum separation between the binary black holes is
$a(1-e)$, the condition (1) is satisfied if
\beq
M_{BH} \ll a(1-e).
\eeq
Then  in terms of the initial comoving separation we have
\beq
y/x \ll 5.8\times 10(x/\bar{x})^{2/3}(M_{BH}/M_{\odot})^{-1/9}(\Omega
h^2)^{-2/9},
\eeq
where we used Eq.(\ref{alpha}) , Eq.(\ref{eccent}) for $\alpha=\beta=1$ and
the relation of $(1-e) \simeq (1-e^2)/2$.
The condition (2) is written as
\beq
R x \ll H^{-1}_{eq}R^2.
\eeq
Then we have
\beq
R \gg 1.0\times 10^{-5}( x/\bar{x})(M_{BH}/M_{\odot})^{1/3}(\Omega h^2)^{2/3}.
\label{kappa}
\eeq
Therefore we have to choose the initial scale factor for the numerical
calculations so that the condition (\ref{kappa}) is satisfied.

\section{three body problem and formation of binary black holes}

We solve  Eq.(\ref{eq:z}) numerically for three body systems using the
fifth-order Runge-Kutta method with the adaptive step size control\cite{recipe}.
Considering the conditions for the validity of the Newtonian
approximation derived in the previous section, we set the initial conditions
 as $x/\bar x = \eta (0.1<\eta<1)$ and $ \dot x=0$
at $R=10^{-3} \eta (M_{BH}/0.5M_{\odot})^{1/3}$
so that  Eq.(\ref{kappa}) is satisfied, where we place a pair of black
holes along the x-axis. 
Note that, using the scaling law of Eq.(\ref{scaling}), 
we see that these initial conditions are the same as
 $x/\bar x = \lambda \eta $ and $ \dot x=0$ at $R=10^{-3} 
(\lambda \eta) (M_{BH}/0.5 \lambda^{-6} M_{\odot})^{1/3}$.
Therefore we can obtain the results for different $M_{BH}$
from a single numerical result.
We then numerically estimate $\alpha$ and $\beta$ for $x$ and $y$ in
the range $0.1 < x/\bar{x} < 1$ and $ 2 < y/x < 7$. 
 The total number of the 
parameters we examined is 100 for each direction of the third body. 
In this section we show the main results  in relation to $\alpha$ and $\beta$ first.
 We will   discuss the  dependence of 
$\alpha$ and $\beta$ on the initial direction of the third body
in Section VI.A. in more detail.
We will also show in  Section VI.D. that the dependence of the results
on the initial conditions is small.

In Fig.1, the trajectories of the second body (the thick curve) and the
third body (the dotted curve) relative to the first body are shown for
(a) $x/\bar{x}=0.3,
y/x=2.0$ and (b) $x/\bar{x}=0.3, y/x=4.0$.
$\theta$ is chosen as $\pi/4$, where $\theta$ denotes 
the angle between the x-direction and 
the direction of the third body.
The coordinate is normalized by $\bar{x}$. 
We see that the binary is formed
through the three body interaction while the third body goes away. 
To see the accuracy of  numerical
calculations, we checked the time reversal of the problem.
 That is,   we have re-started 
the numerical integration from the final
time backward to the initial time.  We have found that the 
differences from the true values of coordinates and velocities 
are very small: the relative error in the coordinate position, $|{\bf z}_{\rm 
init}-{\bf z}({\rm time~ reversed})_{\rm init}|/|{\bf z}_{\rm init}|$, 
is less than $10^{-7}$ and the ``velocity'' component, ${\bf z_i'({\rm time~
reversed})}$, deviates from zero by at most $10^{-5}$.

Fig.2 shows  the semimajor axis $a$ as a function of
initial separation $x/\bar x$. The filled triangles are numerical
results. The solid line is the approximate equation, $a/{\bar
x}=(x/{\bar x})^4$.
We performed the least square fitting of the numerical results assuming that
$a/\bar x=\alpha (x/\bar x)^n$.
It is found that  $a$ is well fitted by the
following function
\beq
{a\over \bar{x}} \simeq 0.41 \left({x\over \bar{x}}\right)^{3.9}
\eeq
irrespective of the direction of the third body as far as we have examined
($\theta=\pi/6,\pi/4,\pi/3$).
The power index $n$ is in good agreement with the analytical estimate in
Eq.(\ref{alpha}) so that we will not discuss the small deviation of $n$ 
from $4$ from now on.

Fig.3 shows $b$/$a$ as a function of $x/y$ 
for $\theta$ equal to (a) $\pi/6$, (b) $\pi/4$ and (c) $\pi/3$.
The filled
triangles are numerical results. The solid line is the approximate
equation $b/a=(x/y)^3$. 
We see that the numerical results are parallel to the approximate estimate
in the previous paper \cite{bhmacho}. 
We performed the least square fitting of the numerical results assuming that
$b/a=\beta (x/y)^n$. The results are given as
\beqa
{b\over a}&=& 0.74 \left({x\over
y}\right)^{3.2}\,(\theta=\pi/6)\nonumber\\
{b\over a}&=& 0.77 \left({x\over
y}\right)^{3.1}\,(\theta=\pi/4)\\
{b\over a}&=& 0.62 \left({x\over
y}\right)^{3.1}\,(\theta=\pi/3)\nonumber
\eeqa
We see  $\theta$-dependence of $\beta$ , which will be discussed in Section
VI.A.  However, as for the power index $n$, it is almost constant so
that we will not discuss the small deviation of $n$ from 3 from now on.

The important conclusion is that 
 we have verified that the power dependence is in good
agreement with the previous analytic order-of-magnitude  
estimate  of Eq.(\ref{alpha}) and
Eq.(\ref{beta}) and that numerical coefficients $\alpha$ and $\beta$
are actually of order unity.
In the next section  we will assume that  $\alpha$ and $\beta$ are
constants. For simplicity we will adopt  $\alpha =0.4$ and $\beta =0.8$.

\section{Probability Distribution Function of  Binaries}

\subsection{Distribution function and binary fraction}

If we assume that black holes are distributed randomly, then the
probability distribution function $P(x,y)$ for the initial comoving
separation of the binary $x$ and the initial comoving separation of the
nearest neighboring black hole from the center of mass of the binary $y$ 
is
\beq
P(x,y)dxdy={9x^2y^2\over \bar{x}^6}e^{-y^3/\bar{x}^3}dxdy,
\label{fae1}
\eeq
where $0<x<y<\infty$ so that
$\int^{\infty}_0 dx \int^{\infty}_{x} dy P(x,y)=1$.
Changing the variables $x$ and  $y$ in Eq.(\ref{fae1})
to $a$ and  $e$ with Eq.(\ref{alpha}) and Eq.(\ref{eccent}),
we obtain the probability distribution function of the eccentricity and
the semimajor axis of binaries as
\beq
f(a,e)dade
= {3\over 4} {\beta \over (\alpha \bar{x})^{3/2}}{a^{1/2}e\over
(1-e^2)^{3/2}}\exp \left[-{\beta \over
(1-e^2)^{1/2}}\left({a\over \alpha \bar{x}}\right)^{3/4}\right]dade,
\label{fae}
\eeq
where $\sqrt{1-\beta^2}<e<1$, $0<a<\infty$ so that
$\int^{\infty}_{0}da \int^{1}_{\sqrt{1-\beta^2}}de f(a,e)=1$.
For $0<e<\sqrt{1-\beta^2}$, $f(a,e)=0$, i.e. no such binary is formed.

Integrating $f(a,e)$ with respect to $e$ in the range
$\sqrt{1-\beta^2}< e<1$, we obtain the distribution
function of the semimajor axis as
\beq
f_a(a)da={3\over 4}\left({a\over \alpha\bar{x}}\right)^{3/4}\exp\left[
-\left({a\over \alpha\bar{x}}\right)^{3/4}\right]{da\over a}.
\label{fa}
\eeq
{}From Eq.~(\ref{fa}), it is found that if $\alpha=1$, 
the fraction of BHMACHOs that are in binaries with  $a\sim
2\times 10^{14}$cm and $M_{BH}=0.5M_{\odot}$ is
$\sim 4\%$ and $\sim 0.4\%$ for
$\Omega h^2=$ 1 and 0.1, respectively.
On the other hand, if we adopt
our numerical estimate of $\alpha$ given in the
previous section,  $\alpha=0.4$, 
the fraction becomes $\sim 7\%$ and $\sim 0.8\%$ for
$\Omega h^2=$ 1 and 0.1, respectively.
This estimated fraction of $\sim 10$ AU size BHMACHO binaries
can be compared with  the observed rate of binary MACHO events (one
binary event in eight observed MACHOs) \cite{bennett}
although  the small number statistics prevents us from stating
 something definite .

\subsection{Gravitational Waves from Coalescing BHMACHO Binaries}

We consider here short period BHMACHO binaries. 
Their coalescence time due to the emission of gravitational waves is
approximately given by\cite{peters}
\beq
 t=t_{0}\left({a\over a_{0}}\right)^4(1-e^2)^{7\over 2}, 
\label{GWt}
\eeq
\beq
 a_{0}=2.0\times 10^{11}\left({M_{BH}\over M_{\odot}}\right)^{3\over
4}\hbox{\rm cm}
\label{defa0}
\eeq
where $t_{0}=10^{10}{\rm year}$ and  $a_{0}$ 
is the semimajor axis of a binary with circular orbit which 
coalesces in $t_{0}$.
Note that Eq.(\ref{GWt}) is an approximation for $e\sim 1$ in
Ref.\cite{peters}.
However it is also a good approximation even for $e\sim 0$.
Eq.~(\ref{GWt}) can be written in terms of $x$ and $y$ using
Eq.~(\ref{alpha}) and Eq.~(\ref{beta}) as
\beqa
t&=&{\bar t} \left({x\over{\bar x}}\right)^{37}
\left({y\over{\bar x}}\right)^{-21},
\label{GWt2}\\
\bar{t}&=&\beta^7\left({\alpha \bar{x}\over a_0}\right)^4t_0.
\label{bart}
\eeqa
Integrating Eq.~(\ref{fae1}) for a given $t$ 
with the aid of Eq.~(\ref{GWt2}), 
we obtain the probability distribution function of the coalescence time  
$f_t(t)$.
We should take the range of the integration as $0<x<\bar{x}$, $x<y<\infty$.
The first condition $x<\bar x$ is necessary for the binary formation.
The second condition  turns out to be $(t/\bar t)^{1/16}\bar x<y<(t/\bar
t)^{-1/21}\bar x$ for a given $t$.
Performing the integration, we have
\beqa
f_t(t)dt&=&{3\over 37}
\left({t\over \bar{t}}\right)^{3/37}\left[\Gamma\left({58\over
      37},\left({t\over{\bar{t}}}\right)^{3/{16}}\right)-
  \Gamma\left({58\over 37},\left({t\over{\bar{t}}}\right)
    ^{-{1/{7}}}\right)\right]{dt\over t}
\nonumber\\
&\cong& {3\over 37}
\left({t\over \bar{t}}\right)^{3/37}\Gamma\left({58\over 
      37}\right){dt\over t},
\label{PDF}
\eeqa
where $\Gamma(x,a)$ is the incomplete gamma function defined by
\be
\Gamma(x,a)=\int^{\infty}_a s^{x-1} e^{-s} ds.
\ee
The second equality is valid when we consider $t\sim  t_0$ because
$t_0/\bar t \ll 1$ for typical values of parameters, that is ,
$t_0/\bar t \sim 2\times 10^{-23}$ for
$\Omega h^2=0.1$, $M_{BH}=0.5 M_{\odot}$, $\alpha=0.4$ and $\beta=0.8$.

  If the halo of our galaxy consists of BHMACHOs of mass $\sim
0.5M_{\odot}$, about $10^{12}$ BHMACHOs exist out to the LMC. The
number of coalescing binary BHMACHOs with $t\sim t_0$ then becomes  
$\sim 1 \times 10^9$  for $\Omega h^2 =0.1$, $\alpha=0.4$ and $\beta=0.8$
so that the event 
rate of coalescing binaries becomes $\sim 1 \times 10^{-1}$
events/year/galaxy.
This rate is slightly larger than the estimate of Ref.\cite{bhmacho}.
On the other hand, 
if the BHMACHOs extend up to half way to M31, the number of 
coalescing binary BHMACHOs with $t\sim t_0$ 
can be $\sim 6 \times 10^9$  and the event rate becomes 
$\sim 6 \times 10^{-1}$ events/year/galaxy.
Both of these estimates are much larger than the best estimate of the
event rate of coalescing neutron stars based on the statistics of binary 
pulsar searches in our Galaxy, $\sim 1\times 10^{-5}$ events/year/galaxy
\cite{phinney,narayan,van}.
Because the first LIGO/VIRGO interferometers in 2001 should be able to 
detect BHMACHO coalescence out to about 15 Mpc distance, i.e., out to the
Virgo Cluster \cite{bhmacho},
the event rate will be several events per year
even if we pessimistically estimate it
($\sim 1/100$ events/year/galaxy in each galaxy like our own).

 In deriving the probability distribution function for the coalescence
 time in Eq.(\ref{PDF}), we have neglected various effects, such as 
the angle dependence of $\beta$, 3-body collision, the effect of the
fourth body, the effect of the mean fluctuation field, the
initial condition dependence and the radiation drag.
 We will consider these effects in the next section.

\subsection{The region checked numerically}

Since we have solved three body problem only for a restricted parameter
range of $x$ and $y$,  one may wonder whether our computations may not
be complete.
Thus we need to show that the parameter range of our calculations is
sufficiently large.

We have verified 
Eq.~(\ref{alpha}) and Eq.~(\ref{beta}) numerically for $x$ and $y$ 
in the range, $0.1 \bar x< x <\bar x$, which means
\beq
10^{-3} < \left({x\over{\bar x}}\right)^3 <1
\label{xrange}
\eeq
and
 $2<y/x<7$, which corresponds to
\beq
8 \left ( {x\over{\bar x}} \right )^3<
\left ( {y\over{\bar x}} \right )^3<
343\left ( {x\over{\bar x}} \right )^3.
\label{check1}
\eeq
On the other hand, $y$ is expressed by  $x$  and  $t$ from  Eq.(\ref{GWt2}).
Therefore if we are interested in the coalescing binaries with the coalescence
time $t_1 < t < t_2$, the range of $y$ is expressed by 
\beq
\left ( {{t_2}\over{\bar t}} \right )^{-{1\over 7}}
\left ( \left({x\over{\bar x}}\right)^3 \right )^{{37}\over{21}}<
\left ( {y\over{\bar x}} \right )^3 <
\left ( {{t_1}\over{\bar t}} \right )^{-{1\over 7}}
\left ( \left({x\over{\bar x}}\right)^3 \right )^{{37}\over{21}}
\label{GWt3}
\eeq
This  range of $y$  determines the probability distribution function
$f_t(t)$ in Eq.(\ref{PDF}) for $t_1 < t < t_2$.
In Fig.4  the horizontal axis and the vertical axis are $(x/{\bar
x})^3$ and $\exp(-(y/{\bar x})^3)$, respectively.
The dashed lines show $x=0.1 \bar x$ and $y/x=$ i (i=$2, 3, 4, 5, 
6, 7$), respectively. Solid lines show  $t_1=0.1 t_0$ and
$t_2=10 t_0$ for $\Omega h^2=0.1$, $M_{BH}=0.5M_{\odot}$, $\alpha=0.4$
and $\beta=0.8$ in Eq.(\ref{GWt3}).
Since the  area in Fig.4 is directly proportional to 
the probability $P(x,y)=d((x/\bar x)^3)d(e^{-(y/\bar x)^3})$ from
Eq.(\ref{fae1}),   almost all the region we are interested 
in ($0.1t_0\siml t \siml 10 t_0$)
 is checked numerically.
So the probability distribution function $f_t(t)$ in Eq.(\ref{PDF}) is
valid for $0.1 t_0\siml t \siml 10 t_0$ though in deriving $f_t(t)$ we
used Eq.(\ref{alpha}) and Eq.(\ref{beta}) for the region of $x$ and $y$ beyond 
our calculations.

\section{CONSIDERATIONS OF VARIOUS EFFECTS}

We shall consider several possible factors which may affect the estimate of 
the  event rate of coalescence.

\subsection{Angle dependence}
So far, we have treated $\beta$ as a constant.
In reality, however, $\beta$ has an angle dependence.
In this subsection we investigate whether the angle dependence of $\beta$
affects the estimate of the event rate. 
In the analytical estimate of Eq.(\ref{beta}),
$b$ is proportional to the tidal force. Since the tidal
force is proportional to  $ \sin(2\theta)$, 
we expect that $\beta$ is also proportional to $\sin(2\theta)$.
In Fig.5 we show the result of numerical calculations for
the angle dependence of $\beta$
averaged for various values of $x$ and $y/x$
for the exponent $n=3$.
We find that  the angle dependence of $\beta$ can be approximated by
\beq
\beta\simeq 0.8 \sin(2\theta).
\eeq

 In the previous section  we have used the
maximum value of $\beta$  so that
if we take this angle dependence into account the effective $\beta$
will decrease.
The probability distribution function $f_t(t)$ is proportional to
$\beta^{-21/37}$ since  $f_t(t) \propto {\bar t}^{-3/37} \propto
\beta^{-21/37}$ in Eq.(\ref{PDF}) and Eq.(\ref{bart}).
Hence qualitatively the effect of the angle dependence is 
to increase the event rate.

If we consider the initial direction of the third body,
the distribution function $P(x,y,\theta)$ for $x$, $y$ and $\theta$
is given by
\beq
P(x,y,\theta)dx dy d\theta=
P(x,y)dxdy{{\sin \theta d\theta}}=
{9x^2y^2\over \bar{x}^6}e^{-y^3/\bar{x}^3}dxdy{{\sin \theta d\theta}},
\label{faeangle}
\eeq
where $0 < \theta < \pi/2$ and 
we assumed that $P(x,y)$ does not depend on $\theta$.
Integrating Eq.~(\ref{faeangle}) for a given $t$ 
with the aid of Eq.~(\ref{GWt2}), we obtain
\beqa
f_t^{ang}(t)dt
&=&{3\over 37}
\int^{{1\over2}\arcsin\left({t/{\bar t}}\right)^{1\over 7}}
_{{{\pi}\over 2}-{{1\over2}\arcsin\left({t/{\bar t}}\right)^{1\over 7}}}
\left({t\over \bar{t}\left(\theta\right)}\right)^{3/37}
\nonumber\\
&&\times\left[
  \Gamma\left({58\over 37},\left({t\over{\bar{t}\left(\theta\right)}}\right)
    ^{{3/{16}}}\right) -
  \Gamma\left({58\over 37},\left({t\over{\bar{t}\left(\theta\right)}}\right)
    ^{-{1/{7}}}\right)\right]
{\sin \theta d\theta}{dt\over t},
\label{PDFangle}\\
\bar{t}\left(\theta\right)
&=&\beta^7\sin^7(2\theta)\left({\alpha \bar{x}\over a_0}\right)^4t_0
=\bar t \sin^7(2\theta),
\eeqa
where $\beta$ is replaced with  $\beta \sin (2\theta)$.
The lower and upper limit of the $\theta$ integral is determined by
$t/\bar t(\theta)=1$ where the integrand
in Eq.(\ref{PDFangle}) becomes 0,
i.e. no binary which survives up to $t\sim t_0$ is produced.
We can integrate Eq.~(\ref{PDFangle}) numerically with respect to
$\theta$ for a given $t$.
For example, if we take $t=t_0$, $M_{BH}=0.5M_{\odot}$, $\Omega
h^2=0.1$, $\alpha=0.4$ and $\beta=0.8$, then $f_t^{ang}(t_0)=1.8\times
10^{-3}/t_0$ while $f_{t}(t_0)=1.0\times 10^{-3}/t_0$.
For $0.1 t_0 \siml t \siml 10t_0$, we can show that
$f_t^{ang}(t_0)>f_{t}(t_0)$.
Therefore the event rate of coalescing binaries may be doubled
if we take into account the angle dependence of $\beta$.

\subsection{3-body collision}
In deriving Eq.(\ref{PDF}),
we take the range of the integration as $0<x<\bar{x},~ x<y<\infty$.
Here we consider the range of $y$ carefully.
If $y<\bar x$, the third body may be bound by the binary in the
radiation dominated era.
If the bound third body collides with the  binary,
a complicated 3-body interaction occurs.
It is a difficult problem to estimate how many binaries whose
coalescence time is $\sim t_0$ are left after the complicated
3-body interaction.
So we shall exclude such a case and estimate the minimum event rate.
Namely, we shall restrict the range of the integration to 
$0<x<\bar{x},~\bar{x}< y <\infty$ \footnote{
  The factor in front of $\bar x$ in the condition $\bar x<y$ may be 
  more than unity for the third body not to be bound in this case, since
  the binary is more massive than a single black hole.
  However we set it unity from now on, since
  the factor is of order $O(1)$ and
  the conclusion that the
  event rate of coalescing binaries is not reduced so much is not
  changed.
  This statement also applies to Eq.(\ref{cond2}).}.
This range turns out to be $\bar x < y < (t/\bar t)^{-1/21}\bar x$
for a given $t$.
Integrating Eq.(\ref{fae1}) for a given $t$ with the aid of
Eq.(\ref{GWt2}), we have 
\beqa
f_t^{3body}(t)dt&=&{3\over 37}
\left({t\over \bar{t}}\right)^{3/37}\left[\Gamma\left({58\over 
      37},1\right) - \Gamma\left({58\over 37},\left({t\over{\bar{t}}}\right)
    ^{-{1\over{7}}}\right)\right]{dt\over t}
\nonumber\\
&\cong&{3\over 37}
\left({t\over \bar{t}}\right)^{3/37}
\Gamma\left({58\over 37},1\right) {dt\over t}.
\label{PDFmin}
\eeqa
The second equality is valid when we consider $t\sim t_0$.
Its ratio to $f_t(t)$ in Eq.~(\ref{PDF}) is
\beq
{{f_t^{3body}(t)}\over{f_t(t)}}={{\Gamma\left({58/ 37},1\right)}\over
  {\Gamma\left({58/ 37}\right)}}\simeq 0.60.
\label{ratio3body}
\eeq
That is, the ratio of the binary formation probability 
without the third body being bound by the binary 
to  the total binary formation probability is about $60\%$ for $t\sim t_0$.
Hence the 3-body collision  reduces the event rate at most 40\%.

If we consider the fourth body, the bound third body may not collide
directly with the binary due to the tidal force of the fourth body to
the third body. The third body may only act as  a satellite of the binary.
If  this fact is taken into account,
the minimum probability distribution function $f_t^{3body}(t)$ in
Eq.(\ref{PDFmin}) increases.
The semimajor axis $a'$ and the semiminor axis $b'$ of the orbit of the
third body will be determined
by the initial comoving separation of the fourth body $z$  as
\beq
a'= \alpha' {{y^4}\over {\bar x^3}}
\eeq
and
\begin{equation}
  b'=\beta' \left({y\over z}\right)^3 a'
\end{equation}
respectively.
In deriving these values, we treat the binary as a point mass
and assume that the analytical estimates of Eq.(\ref{alpha}) and
Eq.(\ref{beta}) are valid for this system.
$\alpha'$ and  $\beta'$ will be different from $\alpha$ and  $\beta$
respectively in this case, since the mass ratio of the total mass of the
binary to the third body's mass is not unity.
However  we set $\alpha'=\alpha$ and  $\beta'=\beta$ from now on
 for simplicity since we are making order estimate of the effect.

The 3-body collision will not occur when $\gamma a<r_{min}'$ is
satisfied,
where $\gamma$ is a constant which takes into account  the uncertainty of the criterion 
for the 3-body collision.
 $r_{min}'$ is the minimum separation between the third body
and the center of mass of the binary and is given by
\beq
r_{min}'=a'(1-e')=a'-\sqrt{a'^2-b'^2} \simg {{b'^2}\over{2 a'}}.
\eeq
The third inequality is almost an equality because the case with $a' \gg
b'$ is considered.
The distribution function of four bodies is given by
\begin{equation}
 P_{x,y,z}(x,y,z)\,dx\, dy\, dz=
  {27\over \bar x^9} e^{-z^3/\bar x^3} 
  x^2 dx\, y^2 dy\, z^2 dz.
\label{xyzDF}  
\end{equation}
To calculate the probability that the third body does not collide
with the binary but is bound to it,
the above distribution function Eq.(\ref{xyzDF})
should be integrated with the constraints
\begin{eqnarray}
 x<y<\bar x,
\label{cond1} \\
 \bar x<z,
\label{cond2} \\
 z^3<\delta {{y^5}\over{x^2}},
 \label{cond3}
\end{eqnarray}
where  $\delta = \sqrt{\beta^2/2\gamma}$ and the last inequality comes 
from  $\gamma a<r_{min}'$.
 The condition (\ref{cond2}) is necessary in order that the fourth body
is not bound in the radiation dominated era.
First, we integrate Eq.(\ref{xyzDF}) with respect to $z$ as
\beq
\int^{\delta{{\left(y/\bar x\right)^5}\over{\left(x/\bar x\right)^2}}}_1
e^{-\left({{z}\over{\bar x}}\right)^3}
d\left({{x^3}\over{\bar x ^3}}\right)
d\left({{y^3}\over{\bar x ^3}}\right)
d\left({{z^3}\over{\bar x ^3}}\right)
=\left[ e^{-1}-e^{-\delta{{\left(y/\bar x\right)^5}\over{\left(x/\bar
          x\right)^2}}} \right] d\left({{x^3}\over{\bar x^3}}\right)
d\left({{y^3}\over{\bar x^3}}\right),
\label{zintegral}
\eeq 
where the range of the $z$ integral is determined by Eq.(\ref{cond2}) and
Eq.(\ref{cond3}).
The first term in Eq.(\ref{zintegral})
can be integrated for a given $t$ with the aid of Eq.(\ref{GWt2}) as
\beqa
\int_{\left[\left({{x}/{\bar x}}\right)^{37}
  \left({{y}/{\bar x}}\right)^{-21}={{t}/{\bar t}}\right]}
e^{-1} d\left({{x^3}\over{\bar x^3}}\right)
d\left({{y^3}\over{\bar x^3}}\right)
&=&\left\{\int^{1}_{\delta^{-111/143}\left({{t}/{\bar t}}\right)^{6/143}}
e^{-1} \left({{y^3}\over{\bar x^3}}\right)^{21/37}
d\left({{y^3}\over{\bar x^3}}\right)\right\}
d\left(\left({{t}\over{\bar t}}\right)^{3/37}\right)
\nonumber\\
&=&{{3}\over{37}} \left({{t}\over{\bar t}}\right)^{3/37}
\left[{{37}\over{58e}}\left(1-\delta^{-174/143}\left({{t}\over{\bar
          t}}\right)^{348/5291}\right)\right]{{dt}\over{t}}.
\eeqa
The upper limit of this integral is determined by Eq.(\ref{cond1}),
and the lower limit is determined
by Eq.(\ref{GWt2}) and $\bar x^3< \delta{{y^5}/{x^2}}$
with Eq.(\ref{cond2}) and Eq.(\ref{cond3}).
The second term in Eq.(\ref{zintegral})
can be integrated in the same way.

On the other hand, the probability distribution function
of the binary formation without the
third body being bound by the binary is already obtained in
Eq.(\ref{PDFmin}).
Then, by summing the case with the third body being bound by the binary
and the case without the third body being bound by the binary,
the probability distribution function without the collision between
the third body and the binary is obtained as
\begin{eqnarray}
  f_t^{4body}(t)dt&=&f_t^{3body}(t)dt
  +{3\over{37}} \left({t\over{\bar t}}\right)^{3/37}
  \Biggm[{{37}\over{58 e}}
    \left(1-\delta^{-174/143}
      \left({t\over{\bar t}}\right)^{{348}/{5291}}\right)
\nonumber\\
    &-&{{111}\over{143}}\delta^{-174/143}
    \left({t\over{\bar t}}\right)^{{348}/{5291}}
    \left\{\Gamma\left({{174}\over{143}},1\right)
      -\Gamma\left({{174}\over{143}}, \delta
  \left({t\over{\bar t}}\right)^{-{{2}\over{37}}}\right)\right\}\Biggm]
  {{dt}\over t}
  \nonumber\\
  &\cong&f_t^{3body}(t)dt
  +f_t(t)
  \left[{{37}\over{58 e \Gamma\left({{58}\over{37}}\right)}}\right]dt
\end{eqnarray}
The relative error in the second equality is about a few \% for $t\sim
t_0$ and $\gamma \sim 1$.
The second term is the probability distribution function
for the third body to be bound by the
binary but not to collide with the binary.
For simplicity, we set $\gamma = 1$,
which corresponds to $\delta \cong 0.57$ for $\beta=0.8$.
The ratio $f_t^{4body}(t_0)/f_t(t_0)$ is $82\%$ for 
$M_{BH}=0.5M_{\odot}$, $\Omega h^2=0.1$, $\alpha=0.4$ and $\beta=0.8$.
Comparing this ratio to the ratio in Eq.(\ref{ratio3body}),
the hierarchical three body bound system may be produced by
about $22\%$ of
the binaries that coalesce at $t\sim t_0$,
during the radiation dominated era.
During the radiation dominated era,
the probability of the 3-body collision is about $18\%$ of
the binaries that coalesce at $t\sim t_0$ 

Because the hierarchical structure consisting of a binary and a
satellite becomes unstable by $t\sim t_0$ if $r'_{min}/a$ is very close
to $1$, we may have to take $\gamma$ larger than unity.
Moreover $\gamma$ will depend on the eccentricity of the binary, the
inclination of the orbital plane of the third body and so on.
Here `unstable' means that the third body crosses the binary and the
complicated three body interaction occurs.
There are some criterions for the stability of the binary
(e.g. Ref.\cite{hierarchy}).
If we need to estimate the event rate accurately,
we will have to pay attention to the value of $\gamma$.
However we do not need such an accuracy here, so we set $\gamma$ as a
constant.
For $\gamma = 10$, the ratio $f_t^{4body}(t_0)/f_t(t_0)$ is $71\%$ for 
$M_{BH}=0.5M_{\odot}$, $\Omega h^2=0.1$, $\alpha=0.4$ and $\beta=0.8$.

In addition to the stability of the hierarchical system,
there may be something to be considered about the value of $\gamma$.
If the third body is not separated enough from the binary,
the tidal force from the third body deforms the orbit of the
binary more effectively than the gravitational wave does.
Then the estimate of the life time of the binary using
Eq.(\ref{GWt}) is not adequate for $t\sim t_0$,
which results in the change of the event rate estimate.
Note that a little change in the eccentricity, $e$, causes a large 
change in the life time in Eq.(\ref{GWt}) when the orbit is very
eccentric, $1-e^2 \ll 1$.
The effect of the orbital deformation will be discussed in Section VII.

\subsection{Effect of mean fluctuation field}
In this subsection, we will estimate the tidal force from bodies other 
than the third body and how it affects the estimate of the event rate.
The tidal field from the $i$-th BH will be given by 
\begin{equation}
 T_i \propto {1\over x^3_i},
\end{equation}
where $x_i$ is the comoving separation of $i$-th BHMACHO from the center
of mass of the binary.
The distribution function for $x_4$, $x_5$, $\cdots$, $x_i$ is
\beq
P(x_4, x_5, \cdots, x_i) = e^{-{x_i^3}/{\bar x^3}}
d\left({{x_4^3}\over{\bar x ^3}}\right)
d\left({{x_5^3}\over{\bar x ^3}}\right)
\cdots
d\left({{x_i^3}\over{\bar x^3}}\right).
\eeq
Then the mean value of $(1/x^3_i)^2$ for given $x_3$ is estimated as
\beqa
\langle {1\over{x_i^6}} \rangle & = &
{{\int_{x_3^3/\bar x^3}^{\infty} d\Big({{x_4^3}\over{\bar x^3}}\Big)
\int_{x_4^3/\bar x^3}^{\infty} d\Big({{x_5^3}\over{\bar x^3}}\Big)
\cdots
\int_{{x_{i-1}^3}/{\bar x^3}}^{\infty} d\Big({{x_i^3}\over{\bar
    x^3}}\Big)
e^{-{x_i^3}/{\bar x^3}}{1\over{x_i^6}}
}\over{
\int_{x_3^3/\bar x^3}^{\infty} d\Big({{x_4^3}\over{\bar x^3}}\Big)
\int_{x_4^3/\bar x^3}^{\infty} d\Big({{x_5^3}\over{\bar x^3}}\Big)
\cdots
\int_{{x_{i-1}^3}/{\bar x^3}}^{\infty} d\Big({{x_i^3}\over{\bar
      x^3}}\Big)
e^{-{x_i^3}/{\bar x^3}}
}}
\nonumber\\
& = & {1\over{\bar x^6}}
\int_f^{\infty} dv_4 \int_{v_4}^{\infty} dv_5 \cdots 
\int_{v_{i-2}}^{\infty} dv_{i-1} 
\int_{v_{i-1}}^{\infty} {dv_i\over v_i^2} e^{-v_i}
\Bigg/ e^{-f},
\eeqa
where 
\begin{equation}
 f:= {x_3^3\over \bar x^3}. 
\end{equation}
Thus the mean value of the tidal field for given $x_3$ 
will be estimated as
\begin{eqnarray}
 \langle T^2 \rangle & = & N \sum_{i=3}^{\infty}
    \langle {1\over x_i^6} \rangle
\cr
 & = & {N\over x_3^6}
   \left[1+f^2 \sum_{i=4}^{\infty} 
          \int_f^{\infty} dv_4 \int_{v_4}^{\infty} dv_5 \cdots 
          \int_{v_{i-2}}^{\infty} dv_{i-1} 
          \int_{v_{i-1}}^{\infty} {dv_i\over v_i^2} e^{-v_i +f} \right].
\end{eqnarray}
The first term is the contribution from the third object and 
the second term is that from the other objects. 

For $i\ge 5$, 
\begin{eqnarray}
 I_i & := & e^{f}\int_f^{\infty} dv_4 \int_{v_4}^{\infty} dv_5 \cdots 
          \int_{v_{i-2}}^{\infty} dv_{i-1} 
          \int_{v_{i-1}}^{\infty} {dv_i\over v_i^2} e^{-v_i}
\cr 
 & = &  e^{f}\int_f^{\infty} dv_4 \int_{v_4}^{\infty} dv_5 \cdots 
          \int_{v_{i-2}}^{\infty} {dv_{i-1}\over v_{i-1}} 
          \int_{1}^{\infty} {dx \over x^2} e^{-v_{i-1} x}
\cr 
 & = &  e^{f}\int_f^{\infty} dv_4 \int_{v_4}^{\infty} dv_5 \cdots 
          \int_{1}^{\infty} {dy \over y} 
          \int_{1}^{\infty} {dx \over x^2} e^{-v_{i-2} x y}
\cr 
 & = &  e^{f} \int_{1}^{\infty} {dy \over y^{i-4}} 
          \int_{1}^{\infty} {dx \over x^{i-3}} e^{-f x y}. 
\end{eqnarray}
Thus
\begin{eqnarray}
 \tilde I & := & \sum_{i=5}^{\infty} I_i
\cr & = &  e^{f} \int_{1}^{\infty} {dy} 
          \int_{1}^{\infty} {dx \over x} {1\over xy-1} e^{-f x y}
\cr & = &  e^{f}\int_{1}^{\infty} {dy}\,y'
          \int_{y}^{\infty} dz {1\over z(z-1)} e^{-f z}
\cr & = & e^{f}\left[
   -\int_1^{\infty} dz {1\over z(z-1)} e^{-fz}
   +\int_1^{\infty} dz {1\over z-1} e^{-fz}\right]
\cr & = & e^f{\rm Ei}(-f).   
\end{eqnarray}

On the other hand, for $i=4$
\begin{eqnarray}
I_4:= 
e^{f} \int_f^{\infty} {dv_4\over v_4^2} e^{-v_4} 
& = &  e^f \left(f^{-1} e^{-f}-
     \int_f^{\infty} {dv\over v} e^{-v}\right) \cr
& = & f^{-1} - e^f {\rm Ei}(-f). 
\end{eqnarray}
Hence
\begin{eqnarray}
 I&:=& \sum_{i=4}^{\infty} 
          \int_f^{\infty} dv_4 \int_{v_4}^{\infty} dv_5 \cdots 
          \int_{v_{i-2}}^{\infty} dv_{i-1} 
          \int_{v_{i-1}}^{\infty} {dv_i\over v_i^2} e^{-v_i +f}
          \nonumber\\
   &=&       I_4
             +\tilde I
   =f^{-1}
\end{eqnarray}
Therefore (tidal force by the fourth, fifth$\ldots$ objects)/
(tidal force by the third object) is estimated as 
$\sim f$.
Note that this averaged value of the tidal force by the fourth,
fifth$\ldots$ objects can be evaluated more easily.
Because the distribution is assumed to be random in space,
the one particle distribution function for the fourth, fifth$\ldots$
objects is given by an uniform distribution in
$x > x_3$ with the averaged density of the BHMACHOs.
Hence $\displaystyle{\sum^{\infty}_{i=4}} \langle {1\over{x^6_i}} \rangle
  ={3\over{4 \pi \bar x^3}} \int^{\infty}_{x_3} {1\over{x^6}}
  4 \pi x^2 dx = {1\over{x_3^3 \bar x ^3}}$.
Of course, this gives the same result as before.

If $y>\bar x$, i.e. $f>1$, the tidal force by the fourth, fifth.... objects
dominates the one by the third body.
In deriving $f_t(t)$ in Eq.(\ref{PDF}) or
$f_t^{3body}(t)$ in Eq.(\ref{PDFmin}),
we should take the effect of the fluctuation field into consideration.
If the tidal force increases, $\beta$ increases effectively,
eventually $f_t(t)$ and $f_t^{3body}(t)$ decrease
because $f_t(t)$ and $f_t^{3body}(t) \propto {\bar t}^{-3/37} \propto
\beta^{-21/37}$.
To estimate how the minimum event rate decreases,
we use
\beq
\langle T^2 \rangle = {N\over {x_3^6}} \left
  [ 1+\left({{x_3}\over{\bar x}}\right)^3\right]
\siml {N\over {x_3^6}}\left( 2\left( {{x_3}\over{\bar x}}\right)^3
\right)
\eeq
as the tidal force for $y>\bar x$.
Assuming that the analytical estimate, $b\simeq ({\rm tidal\
  force}) \times ({\rm free\ fall\ time})^2$, is valid as 
in Eq.(\ref{beta}), we have
\beq
b=\sqrt{2} \beta \left({x\over y}\right)^3 \left({y\over {\bar
      x}}\right)^{3\over 2} a.
\eeq
With this equation and Eq.(\ref{alpha}),
assuming that the form of $a$ in Eq.(\ref{alpha}) is valid,
we write Eq.(\ref{GWt}) in terms of $x$ and $y$ as
\beq
t=2^{7\over 2} \bar t \left( {x\over {\bar x}}\right)^{37}
\left( {y\over{\bar x}} \right)^{-{{21}\over 2}}.
\label{GWtflu}
\eeq
We take  the range of the integration as $0<x<\bar x$, $\bar
x<y<\infty$
which means that we exclude the case that the third body is bound to the 
binary.
Integrating  Eq.(\ref{fae1}) for a given $t$ with the aid of
Eq.(\ref{GWtflu}) in the  range of  $\bar x < y <
2^{1/3} (t/\bar t)^{-2/21} \bar x$, we have
\beqa
f_t^{fluc}(t)dt&=&{3\over 37}
\left({t\over \bar{t}}\right)^{3/37}
2^{-{{21}/{74}}} \left[\Gamma\left({95\over 
      74},1\right) - \Gamma\left({95\over 74},2\left({t\over{\bar{t}}}\right)
    ^{-{2\over{7}}}\right)\right]{dt\over t}
\nonumber\\
&\cong&{3\over 37}
\left({t\over \bar{t}}\right)^{3/37}
2^{-{{21}/{74}}} \Gamma\left({95\over 74},1\right) {dt\over t}.
\label{PDFflu}
\eeqa
The ratio of $f_t^{fluc}(t)$ in Eq.(\ref{PDFflu}) to $f_t(t)$ in Eq.(\ref{PDF}) 
is
\beq
{{f_t^{fluc}(t)}\over{f_t(t)}}
={{2^{-{{21}/{74}}} \Gamma\left({95/ 74},1\right)}\over
  {\Gamma\left({58/ 37}\right)}}
\simeq 0.40.
\eeq
Hence the tidal force from bodies other than the third body
reduces the event rate at most 60\% \footnote{
  In $f^{fluc}_t(t)$, the case where the third body is bound to the
  binary is excluded.
  So we will have to compare it with $f_t^{3body}(t)$ not with $f_t(t)$
  to evaluate only the effect of the mean fluctuation field.
  $f_t^{fluc}(t)/f_t^{3body}(t)=0.67$.
}.

\subsection{Initial condition dependence}
So far we  have assumed that the initial
peculiar velocity is vanishing and the initial scale factor when the bodies
begin to interact is
$R=10^{-3}(x/\bar x)(M_{BH}/0.5M_{\odot})^{1/3}$.
We consider here whether the results crucially depend on the initial
conditions or not .

First, we consider the initial angular momentum.
We have assumed that  the angular momentum of the  binary is only from the
tidal force so that it is given by
\begin{equation}
  J=b \sqrt{{{M_{BH}^3}\over{2a}}}
  = \sqrt{{{\alpha \beta^2}\over{2}}M_{BH}^3 {{x^{10}}\over{y^6 \bar x^3}}},
\label{angmom}  
\end{equation}
where we used Eq.(\ref{alpha}) and Eq.(\ref{beta}).
On the other hand, if the BHMACHO binary has 
a relative velocity $v_f$ at the formation epoch, 
the initial angular momentum will be evaluated as
\begin{equation}
 J_f=M_{BH} R_f x v_f,
\end{equation}
where $R_f$ is given by Eq.(2.1).
We note that $v_f<1$. Otherwise, 
the BHMACHO mass becomes comparable with the radiation 
energy within the volume that the BHMACHO sweeps in 
one Hubble expansion time at the formation epoch.  
In this case  the drag effect due to the radiation field 
will be significant so that  BHMACHO will be decelerated eventually and 
$v_f<1$ after all .
We can now evaluate the ratio of the two angular momenta as
\beqa
\left|{J_f\over J}\right|&\simeq&9 \times 10^{-3} {v_f}
\left({{M_{BH}}\over{M_{\odot}}}\right)^{1\over 6}
\left(\Omega h^2\right)^{1\over3}
    \left({x\over \bar x}\right)^{-4}
    \left({y\over \bar x}\right)^{3}
\nonumber\\
&<&4 \times 10^{-3}
    \left({x\over \bar x}\right)^{-4}
    \left({y\over \bar x}\right)^{3}
\nonumber\\
    &=&4 \times 10^{-3}\left({t\over \bar t}\right)^{-1/7}
    \left({x\over \bar x}\right)^{9/7}
\nonumber\\
    &<&4 \times 10^{-3}\left({t\over \bar t}\right)^{-1/7}.
\eeqa
where we used  Eq.(\ref{rform}), Eq.(\ref{mean}) and  Eq.(\ref{GWt2})
for $M_{BH}=0.5M_{\odot}$, $\Omega h^2 = 0.1$, $\alpha =0.4$ and
$\beta =0.8$. For $t\sim t_0$, 
$|J_f/J|$ can be comparable to $7$ for the extreme case of $v_f\sim 1$
since $t_0/\bar t \sim 2 \times 10^{-23}$.
Because $f_t(t) \propto \beta^{-21/37}$, the event rate becomes 
$30\%$ of Eq.(\ref{PDF})  if the initial peculiar velocity is
 the maximum possible value. Hence we see that the event rate is not
reduced so much even if the initial peculiar  velocity is  extremely large.

Second, we consider whether the initial peculiar velocity
considerably changes 
$\alpha$ and  the formation epoch of the binary.
Consider the case that the initial peculiar velocity is comparable to
unity when $R=R_f\sim 10^{-9}$ in Eq.(\ref{rform})
for $M_{BH}\sim 0.5M_{\odot}$ and $\Omega 
h^2 \sim 0.1$. 
{}From Eq.(\ref{eom:multi}), the peculiar velocity is 
damped to $\sim 10^{-6}(x/\bar x)^{-1}$ by the
time when $R=10^{-3}(x/\bar x)$,
since the interaction between bodies can be neglected during this period.
Therefore it is sufficient to investigate the two body problem assuming
that the peculiar velocity  
$ v_i$ is smaller than $ 10^{-6}(x/\bar x)^{-1}$ at $R=10^{-3}(x/\bar x)$.
The result is shown in Fig.6. 
 Fig.6 shows that $\alpha$ and the formation epoch of the binary
do not crucially depend on the initial peculiar velocity.
When $v_i= -10^{-6} (x/\bar x)^{-1}$, the event rate changes because
$f_t(t) \propto \bar t^{-3/37} \propto \alpha^{-12/37}$.
Since $\alpha$ is about $10\%$ larger, the event rate is only $4\%$ smaller.
Note that the ratio $R \dot x/ \dot R x \sim 10^{-3} v_i (x/\bar
x)^{-1}$ does not depend on $R$.

Finally, we consider whether the initial scale factor changes $\alpha$
and the formation epoch of the binary.
We calculated the two body problem using various initial scale factors.
The results are shown in Fig.7. 
As we can see from Fig.7, $\alpha$ and the formation epoch of the binary
do not strongly depend on the initial scale factor.
Qualitatively the event rate increases if the interaction begins earlier
than $R=10^{-3}x/\bar x$, because $\alpha$ decreases.

\subsection{Radiation drag}
In this subsection we consider  whether the force received from the
background radiation 
is greater than the gravitational force between BHMACHOs or not.
There are two kinds of force received from the background radiation; 
(i) the force from the radiation through which a BHMACHO is
traveling \cite{hogan} and (ii)
the force from the radiation which a BHMACHO deflects, namely the
dynamical friction \cite{chandra,binney}.

First, we estimate the force from the radiation which the BHMACHO sweeps.
The force is estimated as
\beqa
F_{rad}&\sim& ({\rm radiation\ momentum\ density})
\times({\rm cross\ section})
\times({\rm velocity\ of\ the\ BH})
\nonumber\\
&\sim& {{\rho_{eq}}\over{R^4}}\times M_{BH}^2 \times v.
\eeqa
When $R\simeq R_m$,
the ratio of the force from the radiation to the gravitational force
between  BHMACHOs is
\beqa
{{F_{rad}}\over{F_{grav}}}&\simeq&
{{M_{BH}^2 v \rho_{eq} / R_m^4}\over{{{M_{BH}^2}/{(x R_m)^2}}}}
\nonumber\\
&\sim& 10^{-11} \left({{M_{BH}}\over{M_{\odot}}}\right)^{2\over 3}
\left(\Omega h^2\right)^{4\over 3} v \left({x\over{\bar x}}\right)^{-4}
\nonumber\\
&\siml&10^{-13} \left({x\over{\bar x}}\right)^{-4}
\eeqa
where we set $M_{BH}=0.5M_{\odot}$ and $\Omega h^2=0.1$.
Therefore the force from the radiation which the BHMACHO sweeps can be
neglected for $x/\bar x \simg 10^{-3}$. Since our numerical
calculations are performed for $x/\bar x > 0.1$, the radiation drag 
force is negligible.

Next, we investigate whether the dynamical friction can be larger than 
the gravitational force between the BHMACHOs.
When a photon passes by a BHMACHO at a distance $b$,
the photon is deflected by an angle $\theta_d \sim 4M_{BH}/b$.
This deflection changes the momentum direction of the photon.
The momentum of the photon
in the incoming direction changes from $p$ to
$p(1-\cos \theta_d)$.
Thus the momentum of the BHMACHO must be changed.
This momentum exchange causes the dynamical friction.
The force that the BHMACHO receives due to the dynamical friction
can be estimated as
\beq
F_{dyn}=\int ({\rm radiation\ momentum\ density})
\times ({\rm velocity\ of\ the\ BH})
\times (1-\cos\theta _d)
2\pi b db.
\label{dfric}
\eeq
This expression may not be precise relativistically,
but this will be a good approximation when $v \ll 1$.
Assuming $\theta _d$ is small,
\beqa
F_{dyn}&\simeq& \int^{b_{max}}_{b_{min}}
{{\rho_{eq}}\over{R^4}} v \theta_d^2 b db
\nonumber\\
&\simeq& {{\rho_{eq}}\over{R^4}} M_{BH}^2 v \ln \Lambda,
\eeqa
where $\ln \Lambda = \ln b_{max}/b_{min}$ is called the Coulomb
logarithm.
$b_{max} (b_{min})$ is the maximum (minimum) of the impact
parameter. In the cosmological situation the horizon scale sets  the 
natural maximum impact parameter. In the case we are considering 
$b_{min} \simeq M_{BH}$.  Therefore $\ln \Lambda$ cannot be so large,
 and   $F_{dyn}$ is the same order as $F_{rad}$.
Hence the dynamical friction can also be neglected.

\section{life of bhmacho binaries after formation}

Finally, we consider how BHMACHO binaries evolve after the equality time.

First, we consider the 3-body collision during the matter dominated era.
For example, let us consider a galaxy with mass $M_G \sim
10^{12}M_{\odot}$ and a radius $R_G \sim 100 {\rm kpc}$.
Then the density of BHMACHOs $n$ is about $n \sim 1/(10^{19} {\rm cm})^3$.
We estimate the time scale $t_{coll}$ for a BHMACHO binary with the 
semimajor axis $a \sim 2 \times 10^{15} {\rm cm}$
(which corresponds to $x/\bar x \sim 0.4$) to collide with other BHMACHO.
Note that  almost all binaries that contribute to the event rate have
$x/\bar x \siml 0.4$ as we  can see in Fig.4.
If we assume the velocity of the binary $v$ to be the virial
velocity $v \sim \sqrt{M_G/R_G} \sim 100 {\rm km/s}$, $t_{coll}$ is
given by
\beq
t_{coll} \sim {{1}\over{n \sigma v}} \sim 10^{12} {\rm yr} \gg 10^{10}
{\rm yr},
\eeq
where $\sigma \sim \pi a^2 \sim 10^{31} {\rm cm}^2$ is the cross section 
for the 3-body collision. Hence the 3-body collision during the
matter dominated era may be small.

Next, we consider whether the tidal force from the other BHMACHOs alters
the coalescence time of the binary or not.
If the tidal field from the other bodies deforms the orbit of the binary
more effectively than the gravitational wave does,
the life time of the binary in Eq.(\ref{GWt}) may be different.
For example, the binary can not coalesce if the increase of the binding 
energy by the tidal force is greater than the decrease by the
gravitational wave emission.
Since the binaries that contribute to the coalescence rate at present
are highly 
eccentric, $1-e^2 \ll 1$, a little change in the eccentricity, $e$,
causes a large change in the life time in Eq.(\ref{GWt}).
If the coalescence time of the binary in Eq.(\ref{GWt}) is different,
the probability distribution function for the coalescence time $f_t(t)$
in Eq.(\ref{PDF}) is different since we use
Eq.(\ref{GWt}) in deriving $f_t(t)$,
and the event rate of the coalescing BHMACHO binaries may be reduced.

So let us consider the tidal field from the other bodies on 
the orbit of the binary after the equality time.
First, we compare the energy loss rate by the gravitational wave
with the binding energy change rate by the tidal field
from the other bodies.
The average energy loss rate by the gravitational wave \cite{peters}
from a binary with the eccentricity $e$ and the semimajor axis $a$ is
given by
\beq
\left |\dot E^{(GW)} \right|
={{64 M_{BH}^5 \left( 1+ {{73}\over{24}} e^2 + {{37}\over{96}} e^4
    \right)}\over{5 a^5 (1-e^2)^{7/2}}}.
\eeq
On the other hand, the energy change rate by the tidal force from the
other body is at most  given as
\beq
\left| \dot E^{(tidal)} \right|
\sim {\rm (tidal\ force)} \times {\rm (velocity)}
\sim {{M_{BH}^2 a}\over{D^3}} \times {{2a}\over{T_B}},
\label{orderE}
\eeq
where $T_B=2\pi \sqrt{a^3/2M_{BH}}$ is the period of the binary and
$D$ is the distance between the source of the tidal force and the center 
of mass of the binary.
The ratio is
\beqa
\left| \dot E^{(GW)}/\dot E^{(tidal)} \right|
&\sim& \left({{D}\over{2 \times 10^{25}{\rm cm}}}\right)^3
    \left({{x}\over{\bar x}}\right)^{-43}
    \left({{y}\over{\bar x}}\right)^{21}
    \left({{M_{BH}}\over{0.5 M_{\odot}}}\right)^{-11/6}
    \left({{\Omega h^2}\over{0.1}}\right)^{22/3}
\nonumber\\
&\sim& \left({{D}\over{2 \times 10^{25}{\rm cm}}}\right)^3
    \left({{t}\over{\bar t}}\right)^{-1}
    \left({{x}\over{\bar x}}\right)^{-6}
    \left({{M_{BH}}\over{0.5 M_{\odot}}}\right)^{-11/6}
    \left({{\Omega h^2}\over{0.1}}\right)^{22/3}
\nonumber\\
&\sim& \left({{D}\over{1 \times 10^{17}{\rm cm}}}\right)^3
    \left({{x/\bar x}\over{0.4}}\right)^{-6}
    \left({{M_{BH}}\over{0.5 M_{\odot}}}\right)^{-21/6}
    \left({{\Omega h^2}\over{0.1}}\right)^{2},
\eeqa
where we used Eq.(\ref{alpha}) and Eq.(\ref{eccent}) with $\alpha=0.4$,
$\beta=0.8$ and  $e\sim 1$ in the first equality.
We used Eq.(\ref{GWt2}) in the second equality and we set $t=t_0$
in the third equality.
Therefore if $D > 1\times 10^{17}$ cm  the coalescence time based on
Eq.(\ref{GWt}) is a good approximation
for a binary with the semimajor axis $a \siml 2\times10^{15}$ cm,
which corresponds to $x/\bar x \siml 0.4$, since
$\left |\dot E^{(GW)}/\dot E^{(tidal)}\right |>1$.
This condition is satisfied in our case since the mean separation of
BHMACHOs is $\sim 2 \times 10^{17}$ cm at the equality time.

Second, we compare the angular momentum loss rate by the
gravitational wave with that by the tidal field of the other body.
The average angular momentum loss rate by the gravitational wave
\cite{peters} for a binary with eccentricity $e$ and semimajor axis
$a$ is given by
\beq
\left |\dot J^{(GW)}\right|
={{32\sqrt{2}M_{BH}^{9/2}(1+{7\over 8}e^2)}\over{5 a^{7/2} (1-e^2)^2}}.
\eeq 
On the other hand, the average rate of the angular momentum change by
the tidal field is at most given as
\beq
\left |\dot J^{(tidal)}\right|
\sim {\rm (tidal\ force)} \times {\rm (length)}
\sim {{M_{BH}^2 a^2}\over{D^3}}.
\label{orderJ}
\eeq
Their ratio is calculated as
\beqa
\left |\dot J^{(GW)}/\dot J^{(tidal)}\right |
&\sim& \left({{D}\over{2 \times 10^{26} {\rm cm}}}\right)^3
\left({{x}\over{\bar x}}\right)^{-34}
\left({{y}\over{\bar x}}\right)^{12}
\left({{M_{BH}}\over{0.5M_{\odot}}}\right)^{2/3}
\left({{\Omega h^2}\over{0.1}}\right)^{22/3}
\nonumber\\
&\sim& \left({{D}\over{2 \times 10^{26} {\rm cm}}}\right)^3
\left({{t}\over{\bar t}}\right)^{-12/21}
\left({{x}\over{\bar x}}\right)^{-90/7}
\left({{M_{BH}}\over{0.5M_{\odot}}}\right)^{2/3}
\left({{\Omega h^2}\over{0.1}}\right)^{22/3}
\nonumber\\
&\sim& \left({{D}\over{2 \times 10^{20} {\rm cm}}}\right)^3
\left({{{x}/{\bar x}}\over{0.4}}\right)^{-90/7}
\left({{M_{BH}}\over{0.5M_{\odot}}}\right)^{-2/7}
\left({{\Omega h^2}\over{0.1}}\right)^{30/7}.
\eeqa
Therefore if $D > 2\times 10^{20}$ cm , $\left |\dot J^{(GW)}/\dot J^{(tidal)}\right |>1$
for a binary with the semimajor axis $a \siml 2\times10^{15}$ cm,
which corresponds to $x/\bar x \siml 0.4$.
If the scale factor $R$ becomes larger than $\sim 10^{3}$,
the mean separation becomes larger than $2 \times 10^{20}$,
so the tidal force can be neglected.
In the matter dominated era  $R$ is given by $(t/t_{eq})^{2/3}$, so  
$t \sim 10^{9/2} t_{eq}$ at $ R \sim 10^3$,
where $t_{eq}=\sqrt{3/8\pi\rho_{eq}}=\sqrt{3 \bar x^3/8\pi M_{BH}}$
is the equality time.
The increase of the angular momentum $\Delta J$
during $t_{eq} < t < 10^{9/2} t_{eq}$ by the tidal force
can be estimated as
\beq
|\Delta J| \sim \int^{10^{9/2} t_{eq}}_{t_{eq}} \dot J dt
\sim \int^{10^{9/2} t_{eq}}_{t_{eq}}
{{M_{BH}^2 a^2}\over{D^3}} dt
\sim \int^{10^{9/2} t_{eq}}_{t_{eq}}
{{M_{BH}^2 a^2}\over{\bar x^3 (t/t_{eq})^2}} dt
\sim {{M_{BH}^2 a^2 t_{eq}}\over{\bar x^3}}.
\eeq
The ratio of the increase of the angular momentum during $t_{eq} < t <
10^{9/2} t_{eq}$ to  the initial angular
momentum of the binary at the formation  is given by
\beqa
|\Delta J/J| &\sim& {{M_{BH}^2 a^2 t_{eq}/\bar x^3}
  \over{b\sqrt{M_{BH}/2a}}}
= \sqrt{{3 \alpha^3}\over{4\pi\beta^2}}
\left({{x}\over{\bar x}}\right)^3 \left({{y}\over{\bar x}}\right)^3
= \sqrt{{3 \alpha^3}\over{4\pi\beta^2}}
\left({{t}\over{\bar t}}\right)^{-1/7}
\left({{x}\over{\bar x}}\right)^{58/7}
\nonumber\\
&\sim& \left( {{x/\bar x}\over{0.5}} \right)^{58/7}
\left( {{M_{BH}}\over{0.5M_{\odot}}}\right)^{5/21}
\left({{\Omega h^2}\over{0.1}}\right)^{16/21},
\eeqa
where we used Eq.(\ref{alpha}) and Eq.(\ref{beta}) in the second equality.
We used Eq.(\ref{GWt2}) in the third equality and we set
$\alpha=0.4$, $\beta=0.8$ and $t\sim t_0$ in the last equality.
Therefore if $x/\bar x < 0.4$, which corresponds to the most binaries
that coalescence at $t\sim t_0$, as  can be seen from  Fig.4,
the increase of the angular momentum during $t_{eq} < t < 10^{9/2}
t_{eq}$ can be neglected.

To conclude, the tidal force from the other bodies can be
neglected if the binary is separated from the other bodies by
greater than the mean separation.
More detailed calculations taking N-body effects
into account are needed to confirm the above arguments on the effect
of the tidal force of the other body on the evolution of the binary
parameters after formation.
The signs of the change rate of the energy
and the angular momentum in Eq.(\ref{orderE}) and Eq.(\ref{orderJ})
are not certain so that we argued only sufficient conditions.
Moreover if the binding energy of the binary does not change
secularly but periodically under the influence of the tidal field,
Eq.(\ref{orderE}) may be an overestimate.
So the effect of the tidal field may be weaker.

\section{summary}

 In this paper we have discussed black hole binary formation
through three body interactions in the expanding universe.
We have confirmed that the order-of-magnitude argument in
Ref.\cite{bhmacho} is valid up within an error of $\sim$50\%. 
Several effects have been considered.
The effect of the 3-body collision and the mean fluctuation field may
reduce the event rate of the coalescing BHMACHO binaries about half.
On the contrary the angle dependence of the tidal force may increase the
event rate about twice.
The results do not crucially depend on the initial peculiar velocity
of BHMACHOs and the initial scale factor when the BHMACHOs begin to
interact.
The radiation drag does not affect the motion of BHMACHOs.
After all, the probability distribution function for the coalescence
time $f_t(t)$ in Eq.(\ref{PDF}) is  a good estimate.
The error in the event rate estimate can be obtained by considering the
minimum event rate.
The minimum event rate can be estimated as
$[1\times10^{-1} ({\rm original\ estimate\ by\ }f_t(t))]\times
[40\% ({\rm 3body\ collision\ effect}+{\rm mean\ fluctuation\ field\
  effect})]\times
[30\% ({\rm maximum\ initial\ peculiar\ velocity\ effect})]
\sim 1.2 \times 10^{-2}$ events/year/galaxy.
Then the event rate will be $5 \times 10^{-2} \times 2^{\pm 1}$
events/year/galaxy including the 
uncertainty from the various effects in a plain fashion.
This suggests that we can at least expect several events per year within
15 Mpc even when the event rate is minimum,
$1\times 10^{-2}$ events/year/galaxy.
This event rate of coalescing BHMACHO binaries is comparable to or greater
than the upper limit of that of coalescing binary neutron stars \cite{phinney}.
The gravitational wave from such coalescence should be able to be detected
by LIGO/VIRGO/TAMA/GEO network.

We have simplified the real situation to the three body problem, so
that N-body effects have not been fully taken into account. They are (1)
the destruction of the formed binary by the 3-body collision between the 
binary and the infalling body after the equal time, (2)
the deformation of the orbit by the tidal field from the other body,
and so on.
Although these effects have been estimated in Section VII and the event rate
estimate does not seem to be influenced by these effects,
more detailed calculations taking N-body effects into account are needed 
to confirm this conclusion.
It is possible to investigate the N-body effect by N-body numerical
simulations.
However the dynamical range of $a$ is very large
($10^5{\rm cm}<a<10^{16}{\rm cm}$), so we need to perform numerical
simulations with large dynamic range using the biggest supercomputer.
This is an important  challenging numerical problem.

Throughout this paper, we  assumed that all BHMACHOs have the same
mass,
although we have outlined the extension to the unequal mass
case in Appendix A. 
This is based on the assumption of a delta-function type 
density fluctuation at the formation.
Even in this case of the delta-function type density fluctuation,
there is a suggestion 
that in reality the IMF 
of primordial black holes may continue down 
to zero mass limit \cite{niemeyer}.
However, the IMF in this case  has a steep rise proportional to  $ \sim M^3$ 
at the lower mass end and an exponential cut-off near the horizon mass.
Hence the picture of the delta-function-like IMF  seems to be valid. 
However, in the case of a general spectrum of the density fluctuation,
we should consider binaries made from  different mass BHMACHOs.

Although we have assumed that the initial distribution of BHMACHOs is random,
the high density region may have a strong correlation.
Presumably this depends on the black hole formation process or the initial
density perturbation spectrum. 
If a strong correlation existed, more binary BHMACHOs may be formed.
This is also an interesting future problem.

\acknowledgments
We would like to thank professor H. Sato for continuous
encouragement and useful discussions. 
We are also grateful to N. Sugiyama, K. Nakao, R. Nishi,
T. Tsuchiya and D. Ida for useful discussions,
and S.~A.~Hayward for a careful reading of the manuscript.
This work was supported by a
Grant-in-Aid  of the Ministry of Education, Culture, and Sports No.09640351.

\appendix

\section{Extension to Unequal Mass Case}

In this appendix, we estimate the semimajor axis and semiminor axis of 
the BHMACHO binary with unequal mass.
We only give the order-of-magnitude estimate along the line of
Ref.\cite{bhmacho} and Section II.

We describe the mass function of black holes as $F(M)$, which is
normalized as $\int^{\infty}_{0} F(M) dM = 1$.
The average mass of black holes $\bar M_{BH}$ can be obtained as
$\bar M_{BH} = \int^{\infty}_{0} M F(M) dM$.
The mean separation of black holes at the time of matter-radiation
equality is given by $\bar x = (\bar
M_{BH}/\rho_{eq})^{1/3}$,
where we assumed that the average in space is equal to the ensemble
average.
Consider a pair of black holes with masses $M_1$ and $M_2$ and a
comoving separation $x$.
This pair will decouple from the cosmic 
expansion if its mean energy density
$\bar \rho_{BH} = (M_1+M_2)/(2 x^3 R^3)$ becomes larger than the
radiation energy density $\rho_r = \rho_{eq}/R^4$.
In terms of $R$, this condition can be written as
\beq
R > R_m \equiv \left({2 \bar M_{BH}\over
    M_1+M_2}\right)\left({x\over \bar{x}}\right)^3
= \xi\left({{x}\over{\bar x}}\right)^3,
\eeq
where $\xi=2 \bar M_{BH}/(M_1+M_2)$.
The semimajor axis $a$ will be proportional to $xR_m$ and is given by
\beq
a=\tilde \alpha \left({2 \bar M_{BH}\over M_1+M_2}\right){x^4\over
  \bar{x}^3}
= \xi \tilde{\alpha}{{x^4}\over{\bar{x}^3}},
\label{uealpha}
\eeq
where $\tilde \alpha$ is a constant of order $O(1)$.
Consider a black hole with mass $M_3$ in the nearest neighborhood of
the binary. Let its comoving separation from the center of mass of the
binary be $y$. Then the semiminor axis $b$ will be proportional to
(tidal force)$\times$(free fall time)$^2$ and is given by
\beq
b=\tilde{\alpha} \tilde{\beta}
\left({M_3 xR_m\over (yR_m)^3}\right)
\left({(xR_m)^3\over (M_1+M_2)/2}\right)
=\left({2M_3\over M_1+M_2}\right)\tilde{\beta}\left({x\over y}\right)^3 a
=\eta \tilde{\beta} \left({x\over y}\right)^3 a,
\label{uebeta}
\eeq
where $\tilde \beta$ is a constant of order $O(1)$ and
$\eta=2M_3/(M_1+M_2)$.
$\tilde \alpha$ and $\tilde \beta$ may depend on mass.

If we assume that black holes are formed randomly, then the
probability distribution function $P(x,y,M_1,M_2,M_3)$ is
\beq
P(x,y,M_1,M_2,M_3) dx dy dM_1 dM_2 dM_3
={{9x^2y^2}\over{\bar x^6}}e^{-y^3/{\bar x^3}}dx dy
F(M_1) F(M_2) F(M_3) dM_1 dM_2 dM_3,
\label{uePDF}
\eeq
where we assumed that $x$, $y$ do not depend on mass.
Eq.(\ref{GWt}) can be written in terms of $x$ and $y$ using
Eq.(\ref{uealpha}) and Eq.(\ref{uebeta}) as
\beqa
t&=&\tilde t \left({{x}\over{\bar x}}\right)^{37}
\left({{y}\over{\bar y}}\right)^{-21},
\label{ueGWt}
\\
\tilde t &=& (\eta \tilde \beta)^7
\left({{\xi \tilde \alpha \bar x}\over{a_0}}\right)^4 t_0.
\eeqa
Integrating Eq.(\ref{uePDF}) for a given $t$ with the aid of
Eq.(\ref{ueGWt}), we obtain the probability distribution function of the 
coalescence time $f_t^{uneq}(t)$ for the unequal mass case.
We should take the range of the integration as $0<x<\xi^{-1/3}\bar x$,
$x<y<\infty$.
The first condition $x<\xi^{-1/3} \bar x$ is necessary for the binary
formation so that $R_m<1$.
The second condition turns out to be $(t/\tilde t)^{1/16} \bar x
< y < \xi^{-37/63} (t/\tilde t)^{-1/21} \bar x$ for a given $t$.
Performing the integration, we have
\beqa
f_t^{uneq}(t)dt
&=&\int^{\infty}_{0} \int^{\infty}_{0} \int^{\infty}_{0}
{{3}\over{37}}\left({{t}\over{\tilde t}}\right)^{3/37}
\left[\Gamma\left({{58}\over{37}},
    \left({{t}\over{\tilde t}}\right)^{3/16}\right)
  -\Gamma\left({{58}\over{37}},\xi^{-37/21}
    \left({{t}\over{\tilde t}}\right)^{-1/7}\right)\right]{{dt}\over{t}}
\nonumber\\
&\times& F(M_1) F(M_2) F(M_3) dM_1 dM_2 dM_3.
\label{uePDFt}
\eeqa
To integrate the above equation with mass,
we need accurate values of $\tilde \alpha$ and
$\tilde \beta$, as well as assuming the form of the mass fuction $F(M)$.
This is left as our future problem.


\newpage
\vskip 0.3in
\centerline{FIGURE CAPTION}
\vskip 0.05in

\newcounter{fignum}
\begin{list}{Fig.\arabic{fignum}.}{\usecounter{fignum}}

\item
The trajectories of the second body (thick curve) and the third body
(dotted curve) relative to the first body for (a) $x/\bar{x}=0.3, 
y/x=2.0$ and (b) $x/\bar{x}=0.3, y/x=4.0$. $\theta=\pi/4$. 
The coordinate is
normalized by $\bar{x}$. 

\item
The semimajor axis $a$ as a function of
initial separation $x/\bar x$. The filled triangles are numerical
data. The solid line is the approximate equation $a/{\bar x}=(x/{\bar
x})^4$.
\item
$b$(semiminor axis)/$a$(semimajor axis) as a function of $x/y$ 
for (a) $\theta=\pi/6$, (b) $\pi/4$ and (c) $\pi/3$. The filled
triangles are numerical data. The solid line is the approximate
equation $b/a=(x/y)^3$. 

\item
The region we have checked numerically, i.e. $x=0.1\bar x$ in
Eq.(\ref{xrange}) and $y/x=$ i (i=2,3,4,5,6,7) in Eq.(\ref{check1}),
and the region that
corresponds to $0.1 t_0< t< 10t_0$.
The horizontal axis is scaled as $(x/\bar x)^3$ and the vertical axis is 
scaled as $\exp(-(y/\bar x)^3)$ so that the area in the figure is
directly proportional to the probability.
We can see almost all the region we are interested in is within the range
that we have checked numerically for $0.1t_0<t<10t_0$.

\item
The angle dependence of $\beta$ assuming that the functional form of $b$
is as in Eq.(\ref{beta}) is shown.
$\beta$ has an angle dependence as $\beta \propto \sin (2\theta)$.
$\theta$ is the angle between the line that connects the binary and the
line that connects the third body and the center of the binary.

\item
This figure illustrates whether the initial peculiar velocity
crucially changes $\alpha$ or the formation epoch of the binary.
The evolution of the relative distance between the binary with
different initial peculiar velocity at $R_i=10^{-3}(x/\bar x)$ is shown.
The case of $v_i=10^{-6}(x/\bar x)^{-1}$ is the upper one.
The case of $v_i=0$ is the middle one.
The case of $v_i=-10^{-6}(x/\bar x)^{-1}$ is the lower one.
We can see the dependence is week.

\item
This illustrates whether the initial scale factor when the bodies
begin to interact crucially changes $\alpha$ or the formation epoch
of the binary.
The cases of $R_i=10^{-3}(x/\bar x)$, $R_i=10^{-4}(x/\bar x)$ and 
$R_i=10^{-5}(x/\bar x)$ are shown.
We can see the dependence is weak.

\end{list}

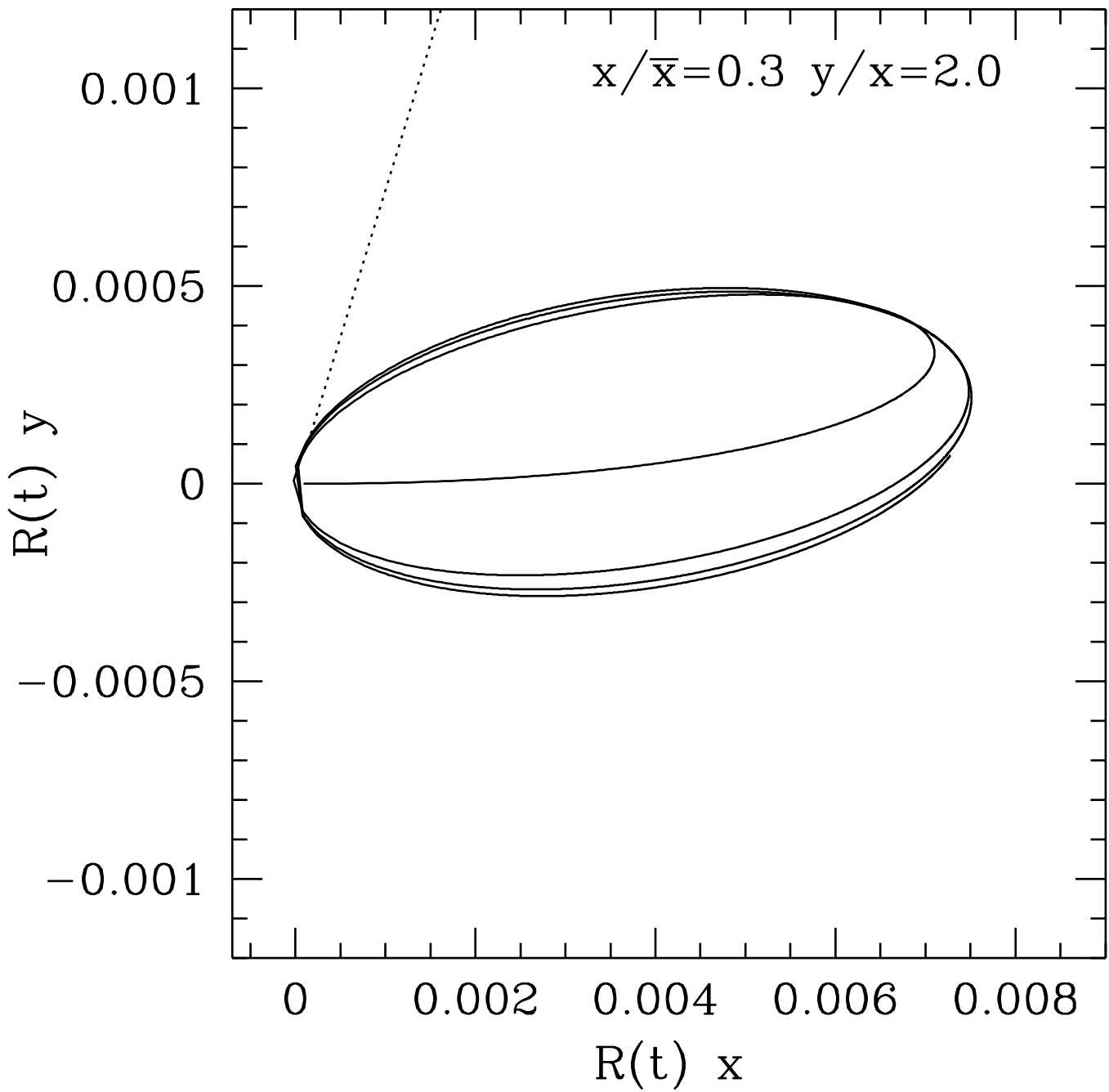
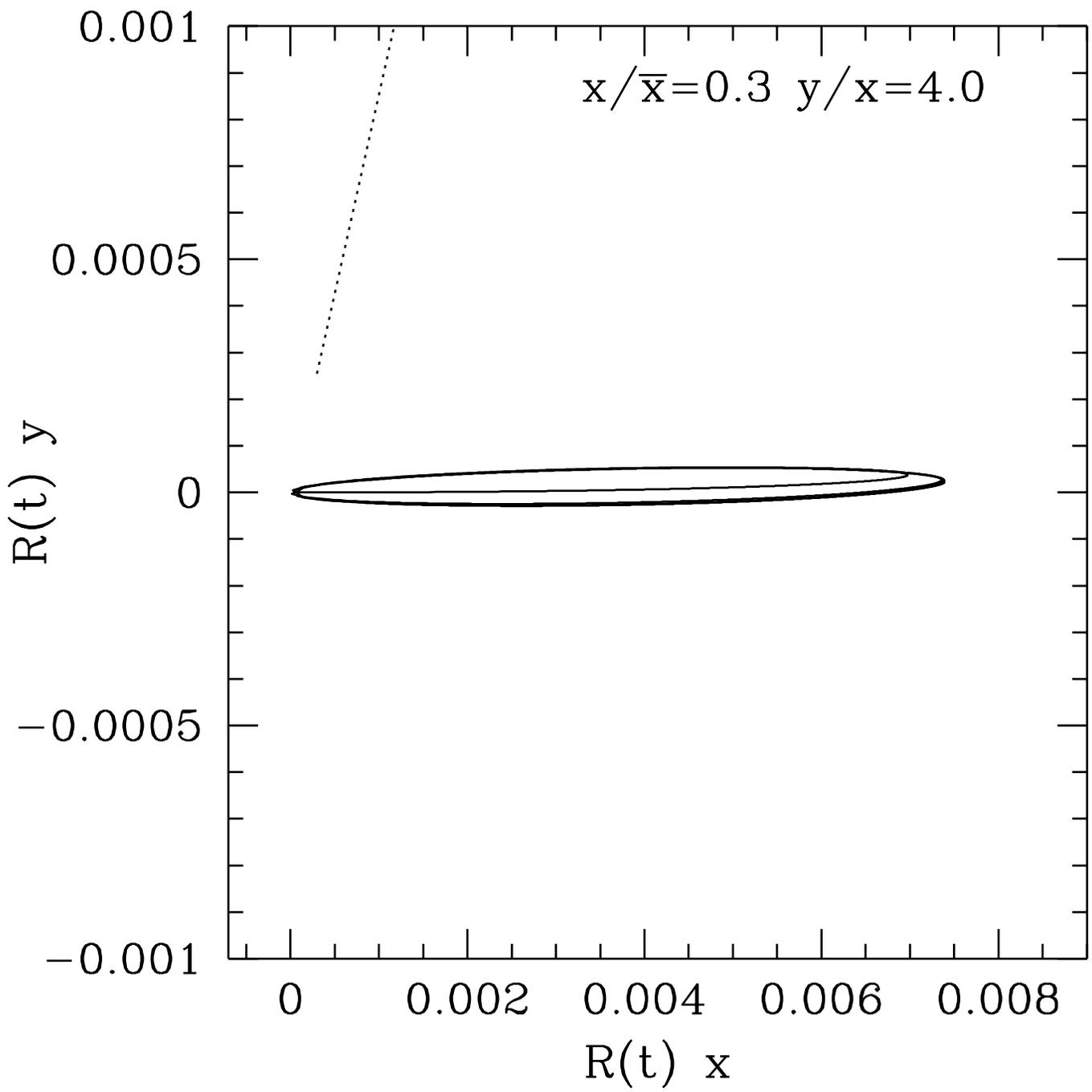
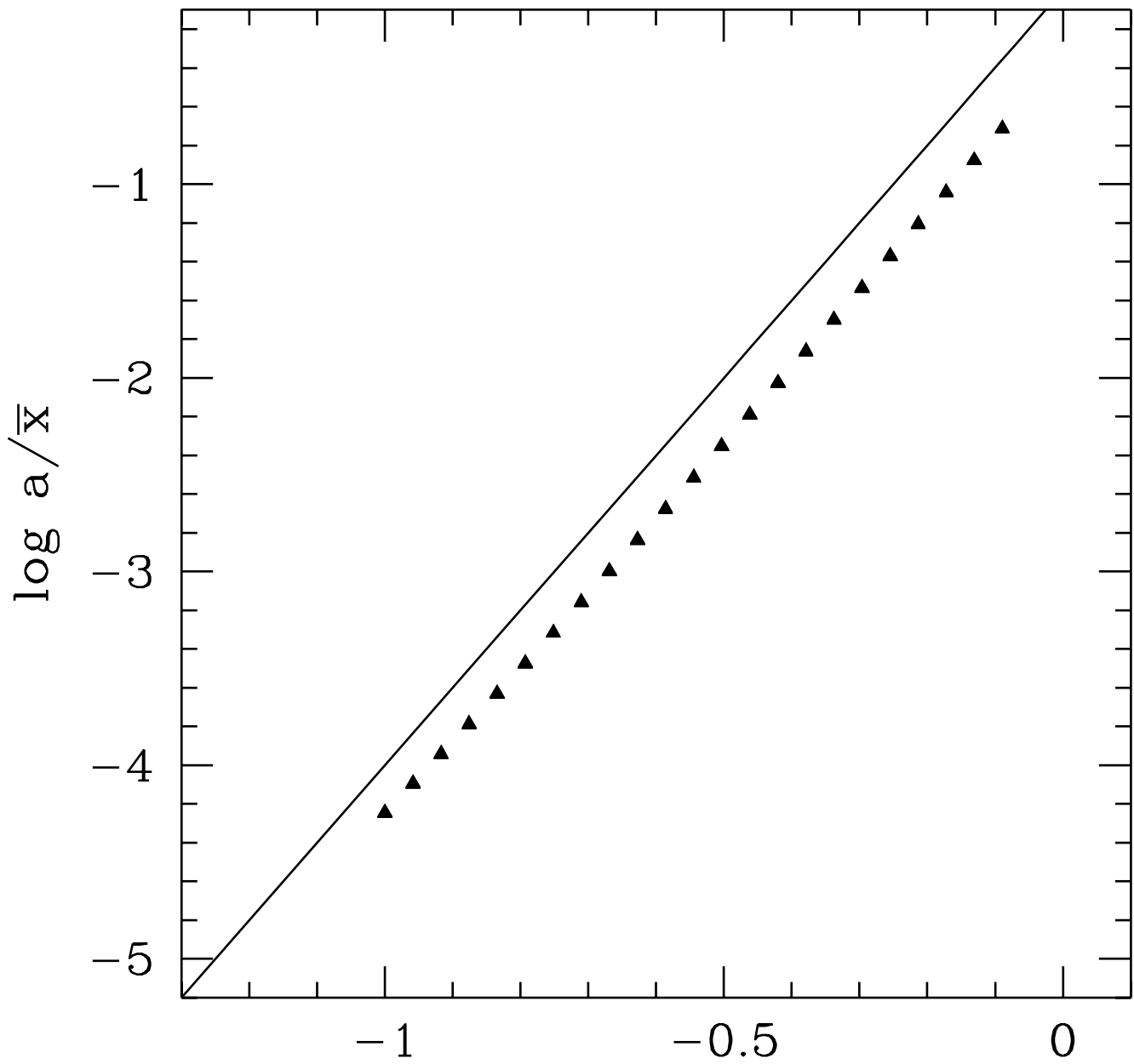
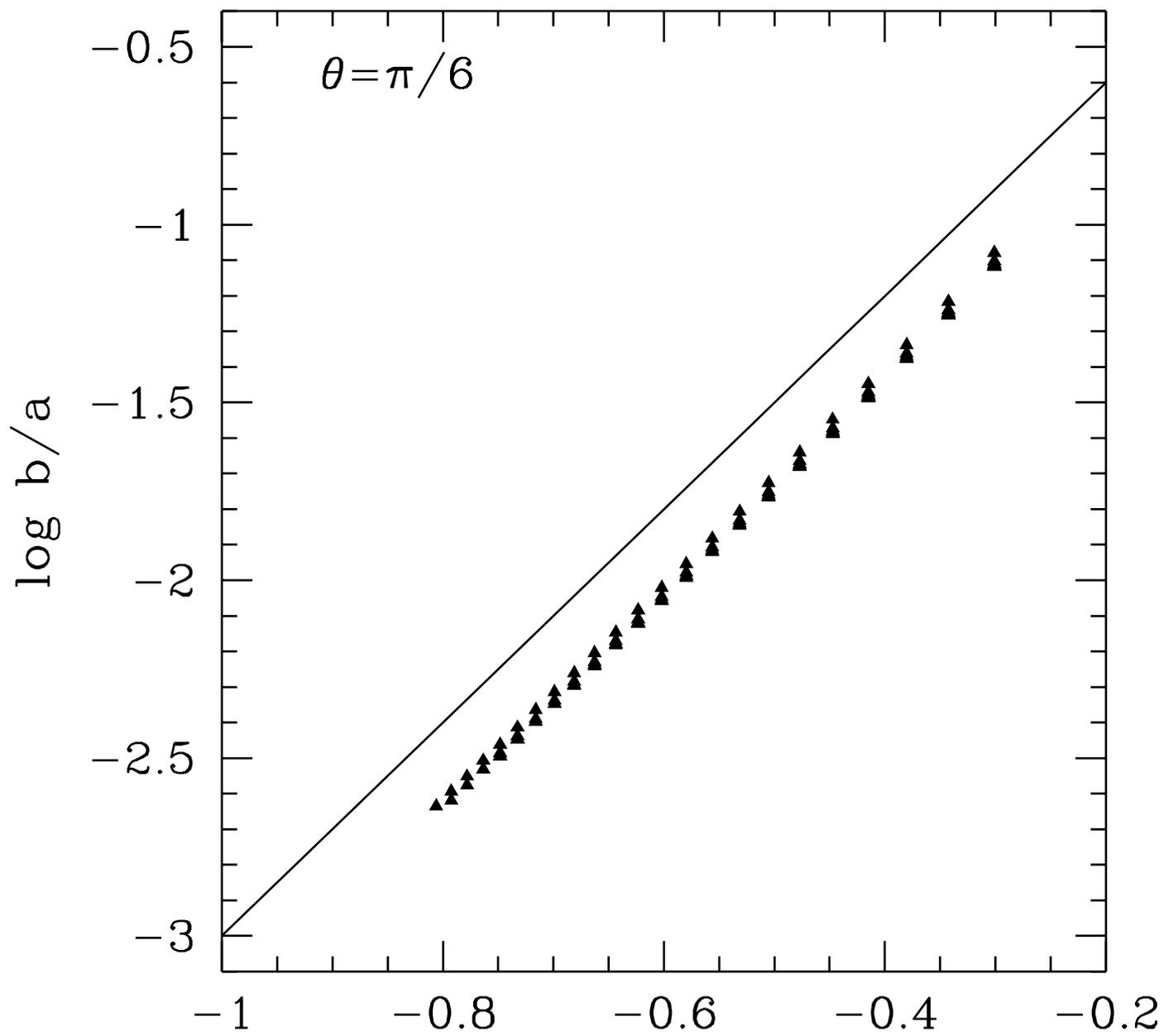
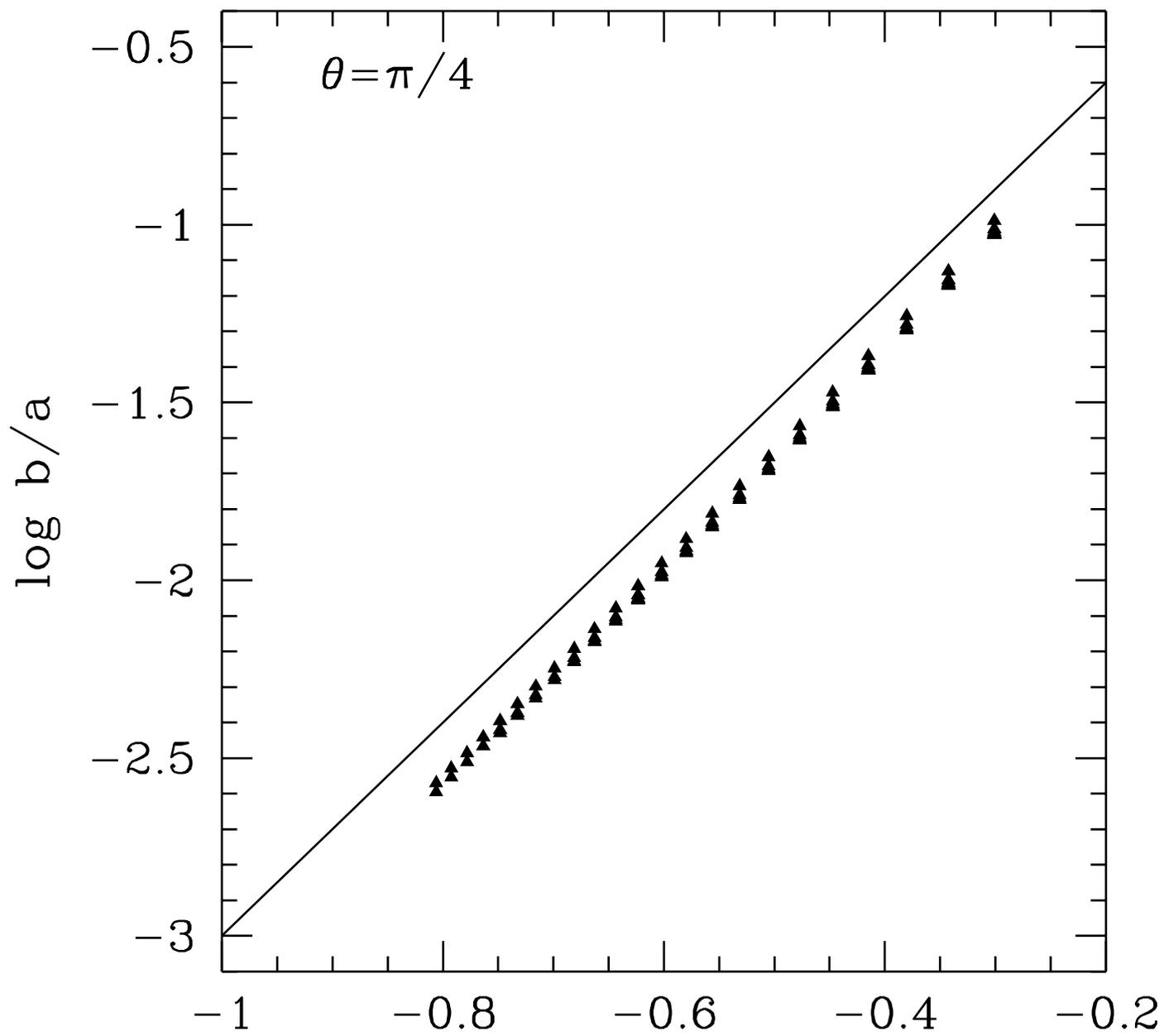
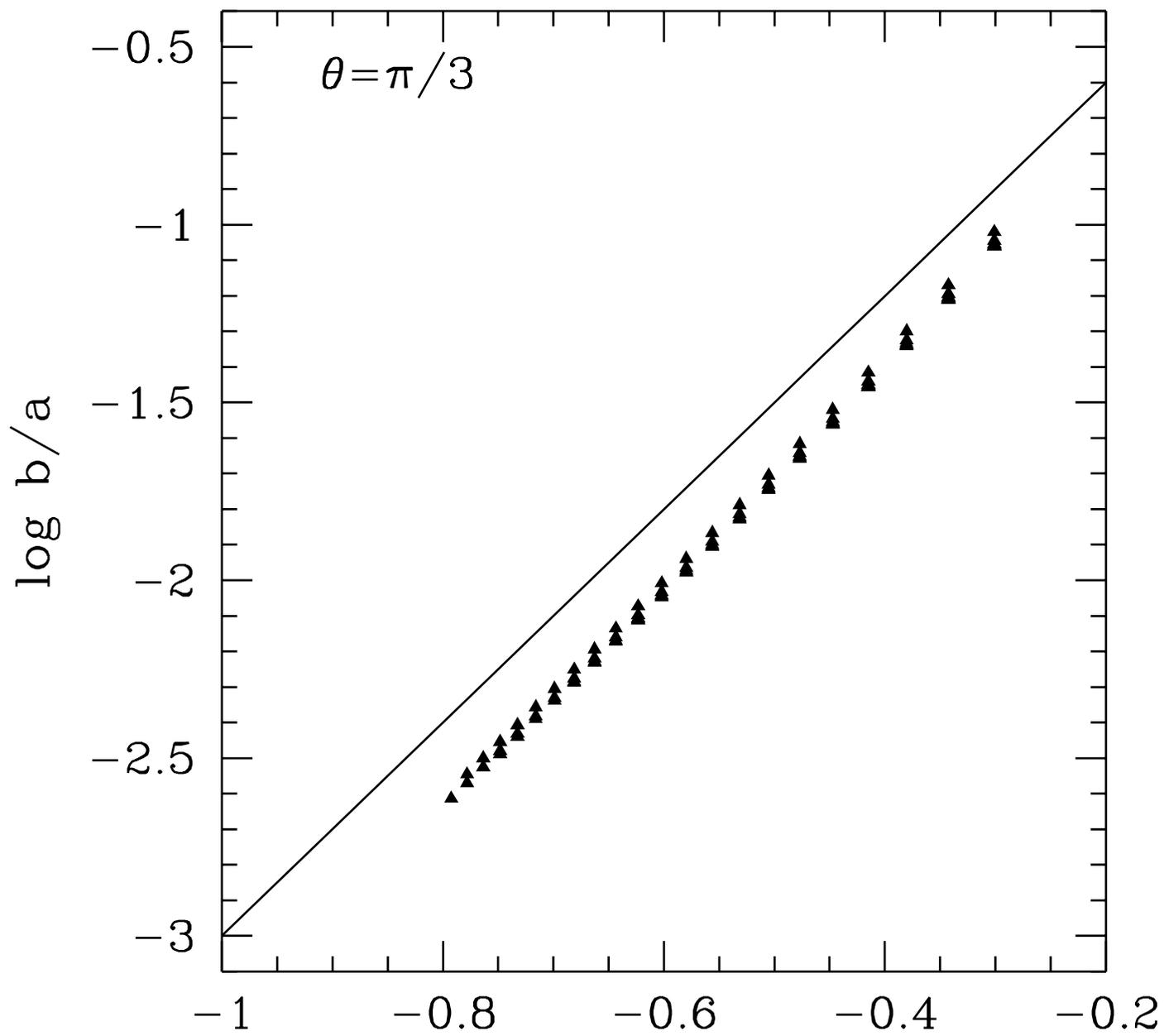
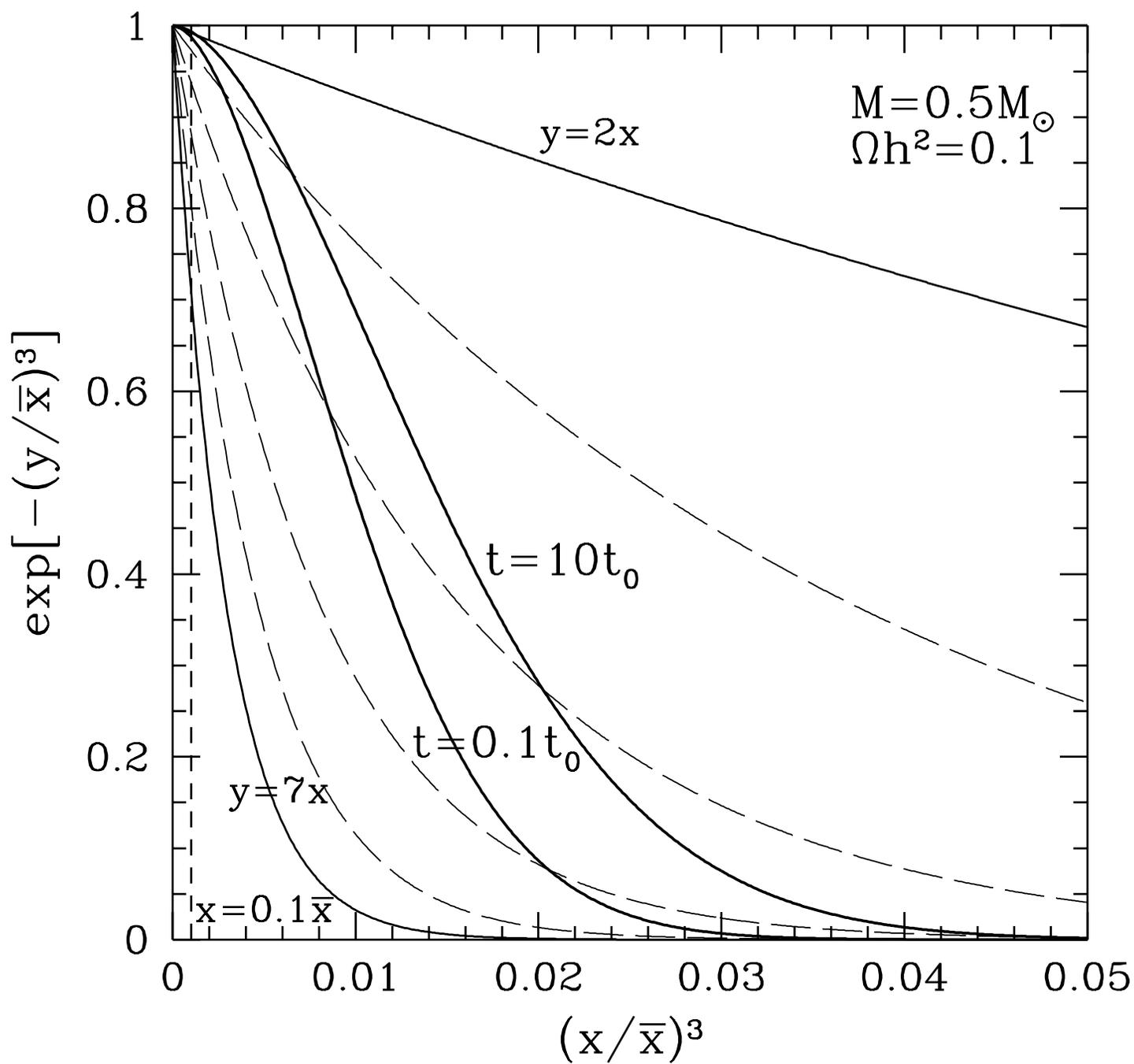
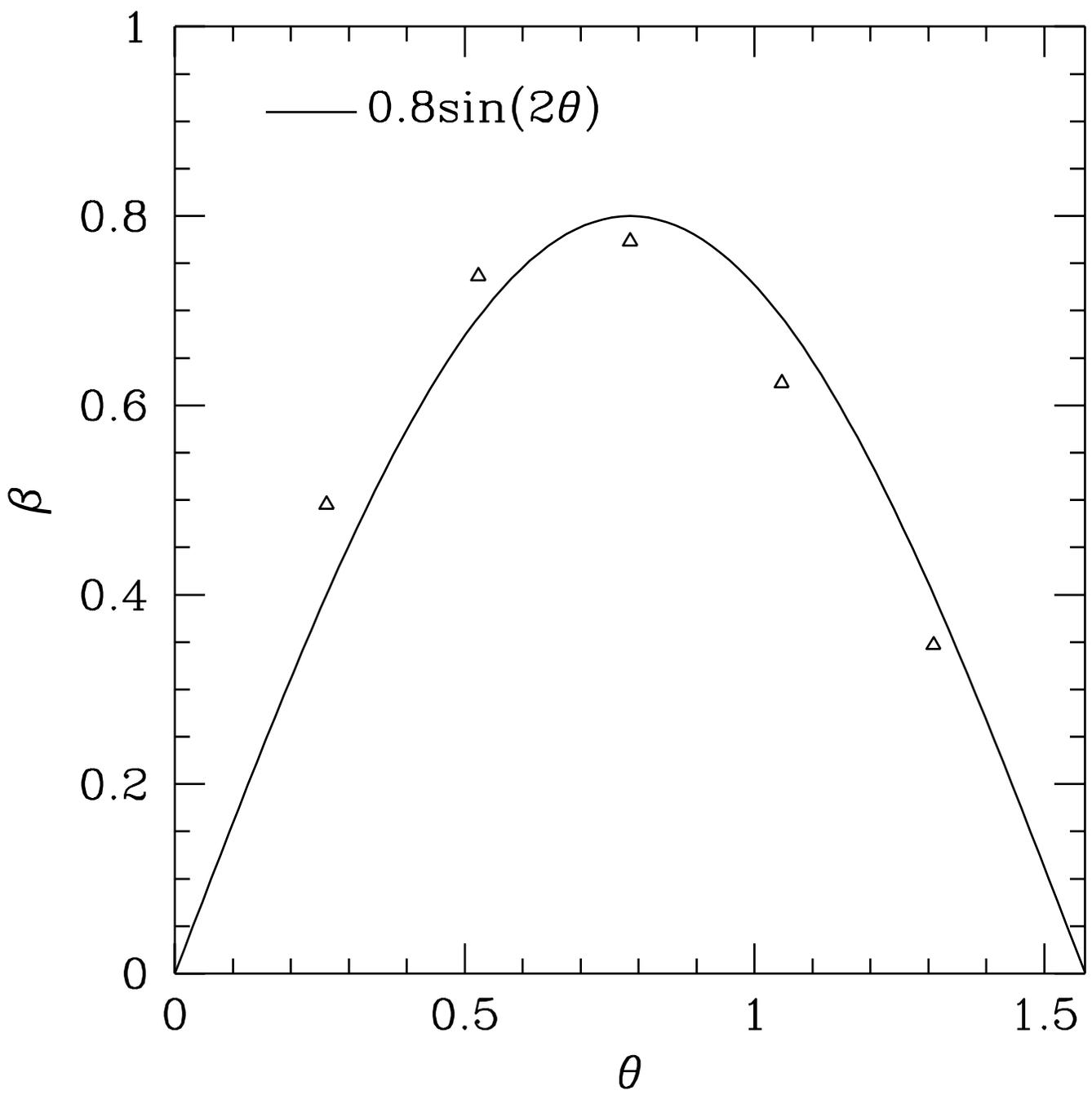
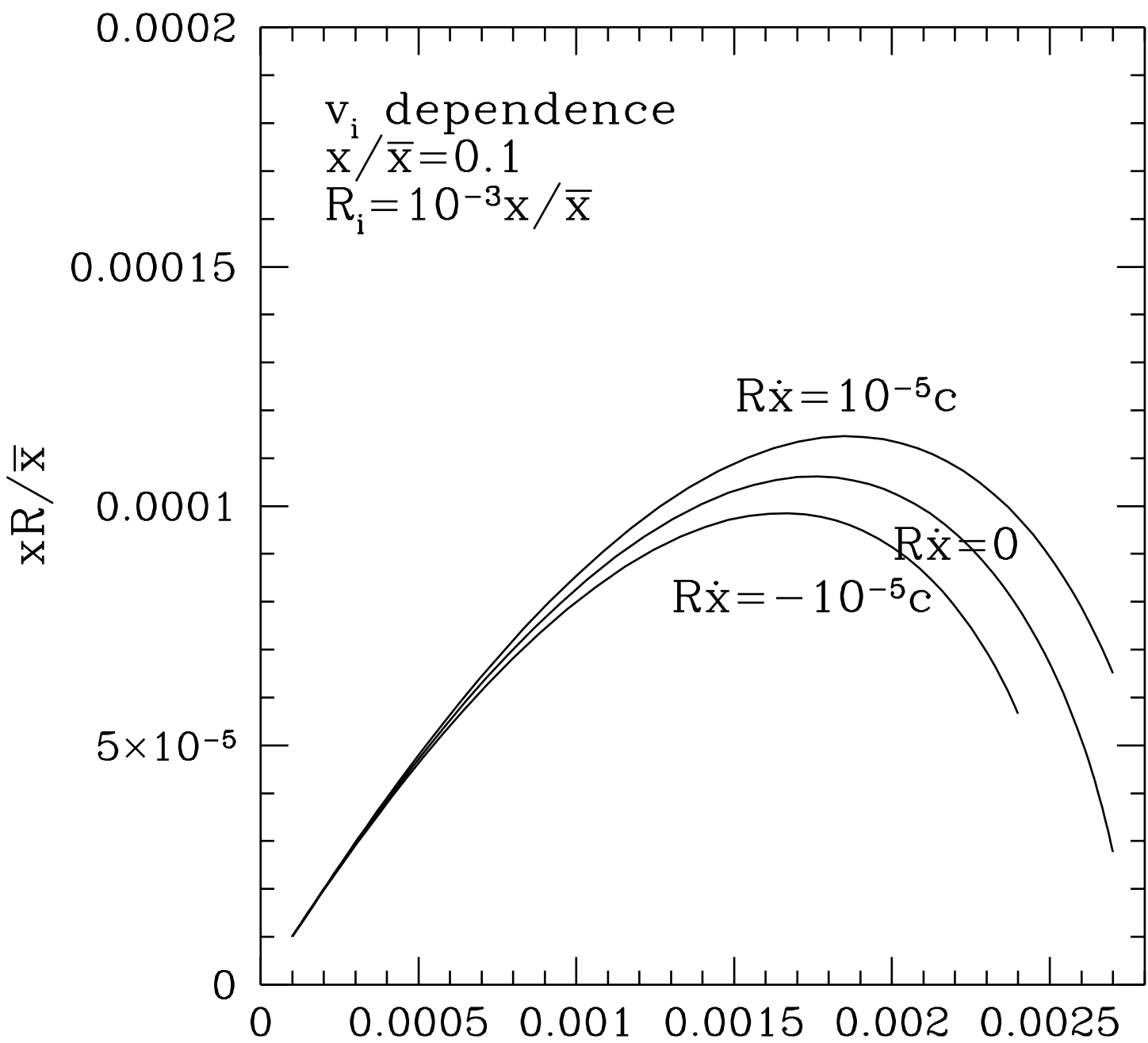
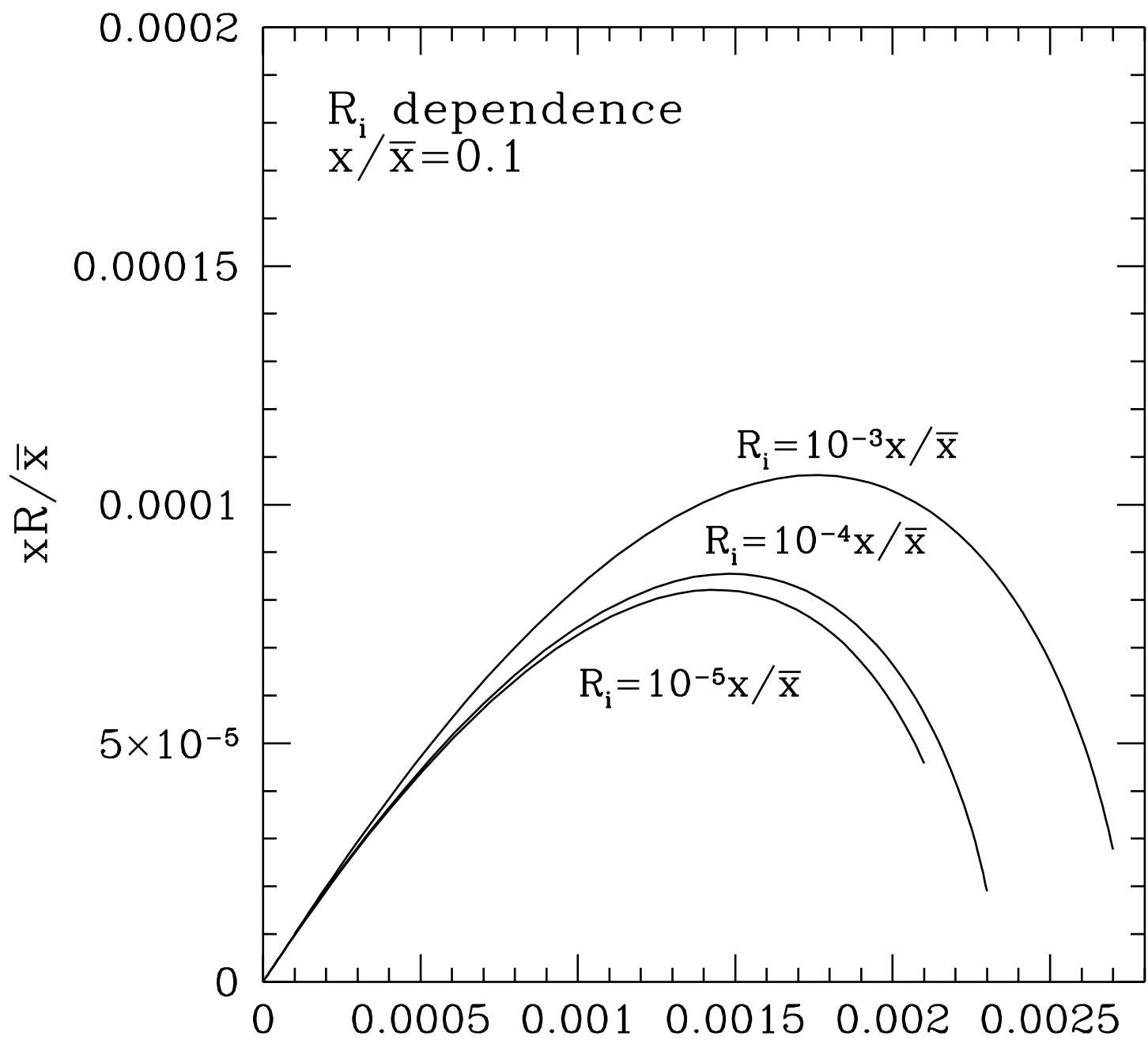
\end{document}